\documentclass[prc,aps,a4paper,groupedaddress,superscriptaddress,
nofootinbib,showpacs,preprintnumbers,onecolumn]{revtex4}

\usepackage{graphicx}
\usepackage{amsfonts}
\usepackage{amssymb}
\usepackage{natbib}
\usepackage{bm}

\begin{document}

\def\ket#1{\vert \,#1 \,\rangle}
\def\bra#1{\langle \,#1 \,\vert}

\def\Kbar{\overline{\!K}}


\title{Low energy $\Kbar N$ interactions and Faddeev calculation
of the $K^- d$ scattering length in isospin and particle bases}

\vspace{0.5cm}

\author{A. Bahaoui}
\affiliation{ 
Universit\'e Chouaib Doukkali, Facult\'e des Sciences,
El Jadida, Morocco }

\author{C. Fayard}
\affiliation{ 
Institut de Physique Nucl\'eaire de Lyon, IN2P3-CNRS, Universit\'e
Claude Bernard,\\ F-69622 Villeurbanne cedex, France }

\author{T. Mizutani}
\affiliation{ 
Department of Physics,
Virginia Polytechnic Institute and State University,\\
Blacksburg, VA 24061, USA}

\author{B. Saghai}
\affiliation{ 
D\'epartement d'Astrophysique, de Physique des Particules, de Physique 
Nucl\'eaire\\ 
et de l'Instumentation Associ\'ee, DSM, CEA/Saclay, 91191 Gif-sur-Yvette, 
France}

\date{\today}

\begin{abstract}
$\Kbar N$ interactions are investigated {\it via} an effective 
non-linear chiral 
meson-baryon Lagrangian. The adjustable parameters are 
determined by a fitting procedure on the $K^-p$ threshold branching ratios 
and total 
cross-section data for $p^{lab}_K\le$ 250 MeV/c. 
We produce predictions for the $\Sigma \pi$ mass spectrum, 
and scattering lenghts $a_{K^-p}$, $a_n(K^-n \to K^-n)$, 
$a_n^0(\Kbar^0 n \to \Kbar^0 n)$, 
and $a_{ex}(K^-p \to \Kbar^0 n)$. 
The $\Kbar N$ amplitudes thus obtained, as well as those for other two-body
channels ($\pi N$, $NN$, and $YN$) are used as input to predict the 
scattering length $A_{K^-d}$,
for which we have devised a relativistic version of the three-body 
Faddeev equations.
Results for all two- and three-body coupled channels are reported both 
in isospin and particle bases. 
All available $\Kbar N$ data are
well reproduced and our best results for
the $K^-p$ and $K^-d$ scattering lenghts are
$a_{K^-p} ~=~ ( -0.90  + i 0.87 )\ \hbox{fm}$ and
$A_{K^-d} ~=~ ( -1.80  + i 1.55 )\ \hbox{fm}$.
\end{abstract}

\pacs{PACS numbers: 11.80.-m, 11.80.Jy, 13.75.-n, 13.75.Jz}
\maketitle

%
\section{Introduction} 
%
The present paper is devoted to the study of the $K^-$-deuteron 
scattering length $A_{K^-d}$ by exploiting a relativistic version of the
three-body Faddeev 
equations in which the principal two-body input is based on  
an effective non-linear chiral meson-baryon Lagrangian in the strangeness 
$S=-1$ sector.  
A preliminary version has been reported in Ref.\cite{Baha02}.  Below we shall 
begin with a survey on the two-body $\Kbar N$ amplitudes 
which are the central input to our present enterprise, as well as 
the evolution of the low energy $K^-d$ physics. In this way we hope that 
our motivation be clear to the reader.      
 
While the low to medium energy {\it kaon-nucleon} processes 
(say, $p_K^{Lab} \le 1.5 $ GeV/c) have been  
known to show no significant structure up to the pion production threshold 
in the $KN$ channel, its $u$-channel counterpart: the $\Kbar N$ channel, 
presents  quite a rich structure (resonances, possible bound states in the 
continuum, etc.):  
see for example \cite{martin,doverwalker}. For  the most recent discussions 
on both experimental and theoretical fronts, see, for example, 
Olin and Park \cite{olinpark}.      
The thus mentioned characteristics may be understood within a simple 
quark model where the quark structure of $K$, $\Kbar$ and $N$ are known 
to be $q\bar s$, $\bar q s$, and $qqq$, respectively with $q$ being $u$ 
and/or $d$ quarks. In this picture the  $\Kbar N$ system may be rearranged 
to become a combination like $(\bar q q)(sqq)$ which 
may be identified, for example,  as $\pi Y,\ (Y=\Lambda, \Sigma)$ in terms 
of the lowest-lying octet hadrons, or to generate strangeness  
$S=-1$ hyperon resonances. 
On the contrary, such a scenario does not materialize for the $KN$ system 
in which the antiquark involved is $\bar s$, hence 
the corresponding low energy process is uniquely the elastic scattering 
of $K$ and $N$. So both theoretically and experimentally, 
the $\Kbar N$ ($S=-1$) system has been drawing much more attention than  
the $KN$ ($S=1$) channel.     
 
Of particular interest in this regard has been the  $K^-p$ channel near 
threshold. It is  
dominated by the below-threshold resonance 
$\Lambda (1405)$ to which it strongly couples.  This resonance decays almost 
exclusively to 
$\pi \Sigma$.  One of the intriguing subjects related to this resonance 
has been on its dynamical origin: whether 
it is a hadronic bound state of $K^-p$ embedded in continuum (since it is 
located above the $\pi \Sigma$ threshold),   
or a three-quark baryon resonance (or something more exotic). Whereas 
no definite 
conclusion  has been drawn based on the 
scattering data analyses by forward dispersion relation 
\cite{martin,doverwalker}, a recent effective  
chiral Lagrangian approach \cite{Ose98} has presented a convincing picture 
in favour of the $K^-p\ (I=0)$ bound state,  
as we shall discuss later in the context of the 
objective of the present paper.  
A more down to the earth, but quite important problem has been the $K^-p$ 
scattering length $a_{K^-p}$, that dictates 
the threshold characteristics of the $K^-p$ interaction.  For this quantity 
the so-called  {\it kaonic hydrogen puzzle}, see e.g. Ref. \cite{Bat90},
had disturbed the community engaged in low energy meson-baryon interactions 
for quite a long  
time. Briefly, the puzzle originated from the fact that the real part of 
$a_{K^-p}$ extracted from the $1s$  
atomic level shift (due to the strong interaction) of the kaonic hydrogen
\cite{oldxray} 
had an opposite sign to the one 
obtained from the analyses of the $K^-p$ scattering amplitude.   
 
Despite the persistence of this puzzle, several pioneering works on the 
low energy negative kaon-deuteron ($K^-d$) scattering by  
solving  three-body equations were performed by simply disregarding 
the information from the  kaonic hydrogen. 
The first study of that type on the $K^-d$ elastic  
scattering  at low energies was performed as early as 1965 \cite{Heth65} 
by adopting simple $S$-wave rank one (non-local) separable interactions 
for the 
$I=0, 1$ \  $\Kbar N$ channels and for the $^3S_1$ deuteron channel as  
basic two-body ingredients.  For  
the $\Kbar N$ interaction the potentials  
were fitted to reproduce the complex-valued $K^-p$ scattering length 
known from the amplitude analysis at that  
time (see for example \cite{dalitztuan}). 
The principal objective of the  
work was to see the convergence properties of the multiple scattering  
series: at low energies they found that the single and double scattering 
contributions were far from sufficient. 
An extension of this work to the three-body break up channels:
$K^-d \to K^- n p, \ \Kbar^0 n n$ was  
performed by the same authors by deforming the momentum integration 
in the complex plane to 
avoid  the anomalous threshold branch cuts due to the final state 
$NN$ interactions \cite{Heth67}. The result turned out to be 
insufficient for 
discriminating the different $K^-p$ input amplitudes by comparing with 
the data, as the two-body input were incomplete, along with insufficient 
statistics for the data. However, the methods developed  
served in motivating  later attempts for studying the three-body final 
states from the theoretical side.    
Also the same authors improved their first elastic scattering calculation
in  Ref. \cite{Heth65} by incorporating the isospin breaking  
effect manifested in  the mass difference between $\Kbar^0$ and $K^-$ as  
well as between $n$ and $p$, which resulted  up to an about $10\%$ difference 
in the cross section \cite{Heth66}.
Later Myhrer \cite{myhrer} studied the role of the  
$\Lambda (1405)$ resonance in the threshold $K^-d$ scattering by assuming 
a simple resonance form for the input  $K^-p$ amplitude. 
Here again the insufficiency of few first iterations of the 
multiple scattering series was found in the presence of a two-body resonance.  
This was then followed  
by using improved $\Kbar N$ amplitudes that took into account  the  
effect of coupling to the physically accessible 
$\pi Y,\ (Y=\Lambda, \Sigma)$ channels and, though implicitly, the effect of 
$\Lambda (1405)$  \cite{Schick}. 
In this regard one should be reminded  that in the 
earliest works mentioned  
above \cite{Heth65,Heth66} those effects were implicitly represented   
only by the complex-valued $\Kbar N$ scattering lengths. To summarize,  all 
the models mentioned here summed up multiple scattering series driven by 
an $S$-wave $\Kbar N$ scattering with a spectator $N$ and  
a $^3S_1$ $NN$ scattering in the presence of a $\Kbar$ spectator. 
We will refer to this type of models as 
{\it Single-Channel Approach}. 
 
The next generation of theoretical endeavour 
\cite{Tok81,Tor86,Bah90,Bah_thesis} may be characterized by explicitly 
taking into account the three-body channels involving hyperons: 
$\pi NY, \ (Y=\Lambda, \Sigma)$, where $Y$ is produced 
from the $\Kbar N \to \pi Y$ reactions in the presence of a spectator 
nucleon.  Also included  
were the two-body interactions $\pi N$ with a spectator $Y$ as well as 
$YN$ with  $\pi$ as a spectator.  In this way genuine 
three-body unitarity was guaranteed to hold.  Common to all three works 
cited here is  the way to construct all the two-body input amplitudes to the 
three-body equations. Apart from the 
$NN$ deuteron channel, all the two-body amplitudes: 
coupled $\Kbar N$-$\pi Y$ channels, $\pi N$, as well as coupled 
$NY$-$NY$  channels, were assumed 
to be obtained from rank one separable 
potentials (mostly in $S$-waves) where the strengths and ranges were adjusted 
to fit the available cross sections, etc.   
Isospin was assumed to be conserved exactly so that the number of channels 
to deal with in the two-body 
input as well as in the three-body equations be contained as manageable. This  
will be termed as the {\it Multi-Channel Approach} as compared with the 
single-channel one mentioned earlier. 
 
The first two of those multi-channel approaches \cite{Tok81,Tor86}  
were dedicated to the  
near threshold break-up reactions: $K^-d \to \pi NY$, where data exist 
for the  
reaction rates, the final mass spectra $m(p\Lambda)$, the neutron momentum 
spectra in the final 
$n\Sigma^- \pi^+$ as well as in $n\Sigma^+ \pi^-$ three-body channels 
\cite{alvarez,tan}. The models were able to reproduce 
the experimental trends depending upon which combinations of signs in the 
non-diagonal amplitudes such as   
$\Kbar N \to \pi \Sigma$, $\Lambda N \to \Sigma N$ should be  adopted.
Note that those amplitudes determined by  
fit to the corresponding cross sections are unique up to an overall sign, 
or phase in more general terms,   
unless additional constraints from, say, 
some symmetries,  were imposed; but  
in three-body processes the sign difference does show up! Also difference 
in the  two-body  amplitudes  
responsible for the final state interaction was found to be visible  
in the break-up channels, so the data  
could be used to discriminate  different two-body models used.  In both works  
the authors also  
calculated the $K^-d$ scattering length $A_{K^-d}$. The best value obtained 
for this quantity in those  
works may be identified as $(-1.34+i\> 1.04)$ fm \cite{Tor86}. 
 
The third multi-channel three-body calculation was carried out in Refs. 
\cite{Bah90,Bah_thesis}  
for low energy elastic $K^-d$ scattering  as well as to find  the best 
theoretical value for $A_{K^-d}$.  
The data used to the $\chi^2$-fits to determine the two-body separable 
interactions were the same as  
in \cite{Tor86}. The novelty was to use relativistic formalism, hence the 
correct kinematics was ensured  
when dealing with different total masses (or different thresholds) in the 
entrance and exit channels,  
both in two-body input amplitudes as well as in  three-body coupled 
equations.   
With the $^3D_1$ partial wave component included in the deuteron channel 
interaction, the best value was  
 $A_{K^-d}= (-1.51+i\> 1.45)$ fm.  
 
After  the last theoretical calculation just discussed  above had come 
out in 1990 \cite{Bah90}, the low energy $\Kbar d$ physics  became dormant 
for about ten years. One may identify some of the possible  reasons 
for that void:  
 
 (i) {\it The kaonic hydrogen puzzle} mentioned earlier kept persisting, 
so it was felt that without any solution to it, one could not find any 
credible low energy $K^-p$ amplitude for use to improve  $K^-d$ models.  
 
(ii) The rank one separable potentials adopted to  model the essential 
ingredients, namely the coupled  $\Kbar N$-$\pi Y$ amplitudes, lacked 
support from underlying strong interaction theory.      
So, even by adopting isospin symmetry to fit the existing data, there was 
no  compelling reason to believe that the {\it best fitting amplitudes} 
be really acceptable on physical ground (a serious attempt of this type
dated back to Henley {\it et al.} and Fink {\it et al.} \cite{henleyetal}). 
In this respect some efforts 
to constrain the fitting parameters by 
$SU(3)$ symmetry deserve to be noted \cite{Sie95,Lee98}. However, 
the separable ansatz still needed to be given a proper justification.  
It should be useful to note that there were also local versions  
of the corresponding potentials based upon 
physical constants from Chiral Lagrangian to be somewhat tuned 
\cite{kaiseretal}, but the local form had no support from the supposedly 
more fundamental theory, either.  
In addition to this, taking into account the isospin breaking effects
would have  pushed  the picture farther into the mist.  
So why should one go forward under such circumstances?  
 
Quite fortunately, there were two major breakthroughs: one on the 
experimental and the other on the theoretical front.   

First, the long haunting {\it kaonic hydrogen puzzle} was finally put to 
an end by the KEK experiment 
\cite{Iwa97}, and  the $K^-p$ scattering length 
$a^c_{K^-p}$, where Coulomb effects are not separated, was extracted to be:  
  $a^c_{K^-p} =      (-0.78 \pm 0.15(stat)  \pm 0.03(syst))  
           + i   (0.49 \pm 0.25(stat)  \pm 0.12(syst))\ \hbox{fm}$,
\noindent 
by applying the Deser-Trueman formula \cite{Deser} to the kaonic hydrogen 
$1s$ level shift $\Gamma$, and width $\epsilon$, by $K^- p$ strong 
interaction : 

$$\epsilon + i \frac{\Gamma}{2} = 2 \alpha^3 \mu^2 a^c_{K^-p},$$ 

\noindent 
with $\alpha$ the fine structure constant and $\mu$ the $K^-p$ 
reduced mass. Although the thus obtained quantity includes the effect of 
the Coulomb interaction, hence not identical to the corresponding quantity due 
exclusively to strong interaction, the difference is, even conservatively, 
at most within a few percent: in the case of pionic hydrogen 
the difference appears to be below $1\%$, see 
for example \cite{deloffpi,Del01}, so the extracted ${\cal R}e(a_{K^-p})$ 
finally was found to have the same sign as the one from the  analysis of 
scattering data. 
The DEAR project \cite{Dear} with the $DA\Phi NE$ facility ($\phi$-factory) 
at Frascati had been  
planning to repeat the experiment with higher precisions, and  
the data taking is reported to have been over.  
In addition, along with other experiments involving various $K$-mesons, 
this project has planned to measure the  
corresponding quantities for the $K^- d$ atom, and the experiment is expected 
to start soon \cite{gianotti}. To obtain the $K^-n$ scattering length 
without recourse to isospin symmetry (hence to find out how good that 
symmetry is realized in the kaon sector in this quantity) is one of the 
objectives for this measurement. But, a more ambitious picture such as 
using the quantities obtained to extract the kaon-nucleon 
Sigma-term: $\sigma_{KN}$, with the help of theories such as 
Chiral Perturbation Theory,  see for example Ref. \cite{ecker}, has been some 
strong driving force for the experiment \cite{olinpark,guaraldo,gensini}.
In fact, such a program has been put into practice recently using 
the pionic hydrogen and deuterium \cite{Sch01}.   
It should be useful to mention here that extracting the 
{\it Coulomb corrected} $K^-d$ scattering length through the  
Deser-Trueman formula may not be as accurate as that for the $K^-p$ system, 
and that relation between the purely  strong and Coulomb corrected $K^-d$ 
scattering length may not be quite simple either.  This was discussed in  
terms of simple models by Barrett and Deloff \cite{Bar99}, 
see also \cite{deloffpi}. This will be revisited in Section \ref{sec:disc} .  
 
Second, based upon chiral perturbation theory,  there has been a steady  
progress in describing the low to medium energy meson-baryon interaction. 
And within the context of our present interest, 
there was an important breakthrough made by Oset and Ramos in describing 
the coupled meson-baryon channel amplitudes for the $S=-1$ 
sector \cite{Ose98}. 
The channels involved are: $\Kbar N$, $\pi Y$, and $\eta Y$.  
The driving (potential) terms to the two-body coupled Bethe-Salpeter 
equations  were taken from the lowest order  
in the effective chiral Lagrangian for the $0^-$ octet mesons 
and $1/2^+$ octet baryons.  
Then, an on-shell ansatz developed in \cite{oller} was introduced, 
which enabled the authors to transform the coupled integral equations 
into a set of algebraic equations where the major task was to deal with 
the integration of various meson-baryon two-particle propagators 
which are ultra-violet divergent. 
With only two parameters  adjusted to very plausible values, 
the cut-off at $p=630$ MeV/c in the momentum integrations to be convergent, 
and  the average octet meson decay constant set as $f=1.15 f_{\pi}$, 
the authors were able to reproduce various low energy data associated with 
the above-mentioned coupled channels impressively  
(the isospin breaking effect in the particle masses was included). 
Particularly, the $\Lambda (1405)$ was generated dynamically as an 
unstable bound state of $K^-p$ in the $I=0$ channel whose decay into 
$\pi Y$ was also correctly reproduced.  As will be discussed in the 
next Section,
what made this approach particularly noteworthy is that the on-shell ansatz 
they adopted turned out to be interpreted as a practical justification 
for a separable representation of
two-body meson-baryon potentials, where 
coupling strengths  are the product of two parts,
one dictated by $SU(3)$ symmetry, the other depending on energy.
Note that this on-shell ansatz was made to be rewritten in a more 
elegant form, viz. in the so-called $N/D$ representation, or cast 
into a once-subtracted dispersion integral representation of the 
two-particle propagators, etc. There, the dynamical left-hand cut 
contributions were shown to be weak, and the divergent integrals 
were made finite by an introduction of several subtraction constants 
adjusted to reproduce the relevant data, including the $K \Xi$ channel
\cite{olleroset,ollermeissner,nievesarriola,garciar,osetramosbennhold}.        
    
Now, upon witnessing the experimental and theoretical progress reviewed 
above, the time is ripe for starting the $K^-d$ scattering study again. 
In fact, the Oset-Ramos result \cite{Ose98} was applied to calculate 
the scattering length $A_{K^-d}$ within what is called the 
fixed center (or fixed scatterer) approximation (FCA) 
by Kamalov et al. \cite{Kam01}. Here we want to claim that a more refined 
approach should  match the Oset-Ramos type amplitudes. More specifically,
we think it necessary to exploit a  reliable method in dealing with 
three-body scattering. 

In this article, we thus present a complete study of the  $K^-d$  
scattering length within the Faddeev equations. 
In Sec. \ref{sec:kbarn_inter} we first adopt the approach due to  
reference \cite{Ose98} and construct  the coupled $s$-wave separable 
$\Kbar N$ interactions, including the coupling to the charge exchange, 
as well as the $\pi Y$ and $\eta Y$  channels.  
We study  different models, with parameters fitted to the available  
experimental data, along with the constraint to remain close to the 
SU(3) values. 
The results are presented both in the {\it isospin} and {\it particle} bases 
where the latter takes into account the isospin breaking effect in terms 
of the physical meson and baryon masses.  
The parametrizations chosen for the deuteron channel are described 
in Sec. \ref{sec:nn_inter}.  Then we discuss other two-body interactions 
adopted in our study in Sec. \ref{sec:other_input}. 
In Section \ref{sec:theory}, we review the structure of the relativistic 
three-body Faddeev equations. In Sec. \ref{sec:results}, we calculate 
the $K^-d$ scattering length by solving those equations,
both in isospin and particle bases. 
To our knowledge, this is the 
first calculation done in the particle basis, permitting to evaluate 
the isospin breaking effects at the three-body level. The sensitivity 
of $A_{K^-d}$ to the two-body input is investigated. 
The discussion is developed in Sec. \ref{sec:disc}, and our conclusions 
are given in the last Section.


\section{Two-body interactions}
\protect\label{sec:two-body}

In this part, we describe the various two-body interactions used as input
in the three-body equations. In the first two Subsections, we present the 
$\Kbar N$ and $N N$ interactions, which are the fundamental ingredients
to the $K^- d$ problem. The last Subsection is devoted to a brief description
of the $\pi N$ and $Y N$ interactions.

\subsection{$\Kbar N$ INTERACTIONS}
\protect\label{sec:kbarn_inter}
%
Here, we intend to construct two-body coupled channel 
$\Kbar N$ interactions.  
The coupled channels involved  are $\Kbar N$, $\pi Y$, and $\eta Y$ 
$(Y=\Lambda, \Sigma)$ with different total charge states for the mesons 
and baryons (in the particle basis), or in different total isospin states 
(in the isospin basis). See Appendix \ref{app-a} for the relation between 
those two representations.     
The physical masses used in the particle basis  and average masses used 
in the isospin basis may be  found  in Table \ref{tab:masses}.  
%
\begin{table}[htb]
\squeezetable
\caption{Particle masses (in MeV). The fourth line gives the average
mass for each isospin multiplet, and the last line specifies the
phase convention used for the isospin states.}
\protect\label{tab:masses}
\begin{center}
\begin{ruledtabular}
\begin{tabular}{cccccccccccc}
  $K^-$       & $\Kbar^\circ$  & $p$        & $n$ &
  $\pi^-$     & $\pi^+$    & $\pi^\circ$ &
  $\Sigma^-$  & $\Sigma^+$ & $\Sigma^\circ$ & $\Lambda$ & $\eta$\\  
\hline 
  493.7  & 497.7 & 938.3  & 939.6 & 
  139.6  & 139.6 & 134.9  &
  1197.4 & 1189.4 & 1192.6 & 1115.7 & 547.4 \\
\multicolumn{2}{c}{$\Kbar$} &\multicolumn{2}{c}{$N$} 
&\multicolumn{3}{c}{$\pi$} &\multicolumn{3}{c}{$\Sigma$}\\ 
%
\multicolumn{2}{c}{495.7}  &\multicolumn{2}{c}{938.9} 
&\multicolumn{3}{c}{138.0} &\multicolumn{3}{c}{1193.1}\\
\hline 
$-\ket{\frac{1}{2} -\frac{1}{2}}$ & $\ket{\frac{1}{2}  \frac{1}{2}}$ &
$ \ket{\frac{1}{2}  \frac{1}{2}}$ & $\ket{\frac{1}{2} -\frac{1}{2}}$ & 
$\ket{1 -1}$ & $-\ket{1 1}$ & $\ket{1 0}$ &
$\ket{1 -1}$ & $-\ket{1 1}$ & $\ket{1 0}$ &
$\ket{0 0}$ & $\ket{1 0}$ \\
\end{tabular}
\end{ruledtabular}
\end{center}
\end{table}
 
    \subsubsection{\bf Separable models}

Let us use $i,\ j,\ k...$, etc. as channel indices. 
Since our present interest is in the $\Kbar p$ near its threshold, 
we may safely assume that 
any given meson-baryon system in the coupled channel is in the relative 
orbital angular 
momentum $s$-state ($\ell=0$). So we may adopt the $s$-wave  projected 
coupled  Bethe-Salpeter 
(or relativistic Lippmann-Schwinger) $t$-matrix equations \cite{Sal51} 
for the transition $j \to i$ which takes the following form:

\begin{equation}
T_{ij} =  V_{ij} + \sum_{k} V_{ik} G^k_0 T_{kj},
\label{eq:tijls}             
\end{equation}

\noindent where $V_{ij}$ is the transition potential, 
and $G^k_0$ is the free meson-baryon 
propagator for the intermediate channel $k$. We note here that 
implicit in the above expression are that (i) the meson-baryon 
systems are in the center of mass frame, and (ii) the integration 
is performed over the off-shell four momentum  associated with channel $k$. 

We take two additional simplifications to make the coupled equations  
manageable. The first one is 
to adopt the Blankenbecler-Sugar procedure to reduce the momentum integration 
from four to three
dimensions \cite{Bla66,Aar77,Gir78}. In particular, the two-particle 
propagator is re-expressed as $G^k_0=G^k_0(p_k;\sigma)$, where $p_k$ is 
the magnitude of the three-dimensional
relative momentum of the intermediate channel $k$, and $\sigma$ is the 
square of the total 
center of mass energy. This may be done  by  taking the discontinuity 
of $G^k_0$ 
over the unitarity branch cut and use it to represent $G^k_0$ in a dispersion
integral form.  We note that this procedure 
results in on  mass shell but off  energy shell form of equations 
which may be regarded as 
a relativistic extension from a familiar non-relativistic
scattering theory.  The second simplification 
step is to assume that the $s$-wave potentials take a 
non-local separable form:

\begin{equation}
V^I_{ij}=g_i(p_i)\lambda^I_{ij}(\sigma)g_j(p_j), \quad I=0,1 \quad, 
\label{eq:separable}
\end{equation}

\noindent where $I$ is the total isospin for the meson-baryon system, and
$g_i$ is the cut-off form factor for channel $i$ which 
is assumed to be a function 
of the magnitude of the three dimensional relative momentum vector in 
the same channel. 
In general the coupling strength $\lambda^I_{ij}$ is assumed to be a 
function of $\sigma$ as 
indicated in the above equation with no {\it left-hand cut} assumed. 
Some rudiments of how our 
coupled two-body $t$-matrices may be obtained with the separable interactions 
is found in Appendix \ref{app-b}. 

Now, when we compare the expression for the coupled $\Kbar N$ channel 
$t$-matrices, Eq. (20) in \cite{Ose98} with our coupled channel $t$-matrices, 
$T_{ij}$ Eq. (\ref{eq:tij-sep}) for 
which the related quantities are defined in Eqs. (\ref{eq:rs}, \ref{eq:gijs}), 
we see immediately that
the two results are identical provided that (i) we set $g_i(p_i)\equiv 1$ 
for all $i$, 
(ii) impose a momentum cut-off $p_{max}$ in the integration in
Eq. (\ref{eq:gijs}), and (iii) set 

\begin{equation}
\label{eq:lamij-os}
\lambda^I_{ij}\equiv-C^I_{ij}\frac{1}{4f^2}(\epsilon_i+\epsilon_j).
\end{equation}

\noindent 
The expression for $\lambda^I_{ij}$ above is from Oset and Ramos \cite{Ose98}, 
which was obtained from the lowest order expansion in $1/f$ of the chiral 
Lagrangian for the octet $0^-$ mesons coupled to the octet ${1/2}^+$ 
baryons.
The  coefficients $C^I_{ij}$ are  due to $SU(3)$ symmetry and tabulated 
as: $C^{I=0}_{ij}=D_{ij},\ C^{I=1}_{ij}=F_{ij}$ in that publication.
These are convenient for the isospin basis, but may be
trivially transformed to the corresponding coefficients 
$C_{ij}$ (for $K^-p$) and $\widetilde C_{ij}$ (for the $K^-n$ 
related channels), for use in particle basis, also tabulated 
in \cite{Ose98}. The corresponding change
to obtain the strength parameters in the particle basis in terms
of the $\lambda^I_{ij}$ follows trivially, see for example Eq. (\ref{eq:a2}).
In our equation  (\ref{eq:lamij-os}) above, $\epsilon_i$ and $\epsilon_j$ 
are the meson energies in the centre of mass system for
the $i$ and $j$ channels, respectively.  

Though  our argument above went just in the opposite direction to  what 
one finds in Oset and Ramos
\cite{Ose98} (see also a  more formally trimmed version of the Oset-Ramos 
line of derivation 
by Nieves and Arriola \cite{nievesarriola}), we have established a 
practical equivalence between the 
separable potential  and the on-shell ansatz for the coupled 
meson-baryon equations. So
within the framework of effective meson-bayon field theory, it is now 
possible to claim that 
the separable ansatz is a very reasonable starting point in describing 
the $s$-wave interactions at low energies.  
In our present work we choose to retain the form factors rather than 
imposing a sharp cut-off. 
This is due to the fact that when solving the three-body equations we 
rotate the momentum 
integration path off the real axis and into the complex plane. For that 
purpose, a sharp cut off is 
not practical. Then in order to respect $SU(3)$ symmetry, we choose 
to use a single form 
factor for all the different channels with a monopole form:

\begin{equation}
\label{eq:gp}
g(p)=\frac{\beta^2}{p^2+\beta^2} ,
\end{equation}  

\noindent
where $\beta$ is the effective cut-off 
momentum.  
Based upon the discussion above we adopt two slightly different types of 
interactions. The first one which we call  OS1 
is just $V^I_{ij}$, Eq. (\ref{eq:separable}), with 
$g_i=g_j\equiv g(p)$, and $\lambda^I_{ij}$ from Eq. (\ref{eq:lamij-os}).
We expect this interaction to produce a  very similar result to the original  
Oset and Ramos \cite{Ose98} model.
The second interaction model is called  OSA, which is a variant of OS1 
in that it
incorporates the possible $SU(3)$ breaking effect in the coupling strengths 
(or in the meson 
decay constant $f$) in terms of extra parameters $b^I_{ij}$
with the substitution: $C^I_{ij} \to b^I_{ij}C^I_{ij}$. 
We then adjust on the relevant data $\beta$, $f$, and $b^I_{ij}$'s, 
the last ones only for OSA, with a constraint that the $SU(3)$ 
breaking effect is reasonably contained. 

It is important to stress here that we follow the 
observation by Oset and Ramos \cite{Ose98} and retain the $\eta Y$ channels 
in our fit although this 
channel has a substantially higher threshold as compared with that for 
$\Kbar N$, see  
Table \ref{tab:masses}.  The necessity for the inclusion of these channels 
will be demonstrated later.
 As a result we have three coupled channels:
 $\Kbar N,\ \pi \Sigma,\ \eta \Lambda$ to deal with for $I=0$, 
and four coupled channels:
$\Kbar N,\ \pi \Sigma,\ \pi \Lambda,\ \eta \Sigma$ for $I=1$.  
In terms of physical 
(or particle) channels the following two groups are separately coupled:

\begin{equation}
\label{eq:ch1-8}
K^-~p~\to~
K^-p~,~ \Kbar^\circ n~, ~\Lambda \pi^\circ~,~ \Sigma^+ 
\pi^-~,~\Sigma^\circ \pi^\circ~,
~\Sigma^-\pi^+~, ~ \Lambda\eta~, ~ \Sigma^\circ\eta,
\end{equation}

\begin{equation}
\label{eq:ch1-5}
K^- n \to K^- 
n~,~\Lambda\pi^-~,~\Sigma^\circ\pi^-~,~\Sigma^-\pi^\circ~,~\Sigma^-\eta.
\end{equation}

\noindent Below in the next Subsection we briefly review available 
experimental data to which the model interactions are fitted and compared. 
                \subsubsection{\bf Experimental data}
Since there are no data associated with the initial $K^-n$ channels 
in Eq. (\ref{eq:ch1-5}), we only discuss the ones in 
Eq. (\ref{eq:ch1-8}). 
Furthermore, in the low energy range of our current interest, 
{\it i.e.} $p_K^{lab} \le 250$ MeV, 
the last two channels involving the $\eta$ meson are physically closed 
as their thresholds are 
substantially higher than the rest. Otherwise the remaining physically 
accessible  
coupled channels are now
strongly influenced by the $I=0$ $\Lambda (1405)$ resonance below 
the $K^-p$ threshold which decays almost exclusively to $\pi \Sigma$. 
Note also that while the $\Kbar^\circ n$ has a slightly higher threshold, 
all the $\pi Y$ channels have lower threshold than that for $K^-p$. 
So altogether there is a very rich structure in this coupled set of channels  
\footnote{Notice that the contributions from the 
$K^- p \to K^\circ \Xi^\circ,~K^+ \Xi^-$ channels, which couple only 
indirectly
to the $\Kbar N$ system, were found insignificant~\cite{Ose98} 
at low energies, and hence are not included in our formalism. 
Moreover, our aim in this work being to use 
the elementary reactions as input in the $K^- d$ elastic scattering, 
the electromagnetic processes 
$K^-~p~\rightarrow~\Lambda  \gamma~,~\Sigma^\circ \gamma$, studied by other
authors~\cite{Sie95,Lee98}, are not considered here.}
..
In the energy range considered here, some 90 data points are available
\cite{Abr65,Cse65,Sak65,Kit66,Kim67,Mast76,Bang81,Cibo82,Evans83}.
These data, obtained between 1965 and 1983, bear unequal accuracies, 
as briefly discussed later.

Moreover, accurate data~\cite{Sak65,Kit66,Hum62,Tov71,Now78} for threshold 
branching ratios are also availabe, {\it i.e.},
\begin{eqnarray}
\label{eq:Br}
\gamma & = & \displaystyle\lim_{k \to 0} 
      {{\Gamma(K^-p \rightarrow \pi^+ \Sigma^-)} \over
       {\Gamma(K^-p \rightarrow \pi^- \Sigma^+)}} = 2.36 \pm 0.04 ,\\    
R_c & = & \displaystyle\lim_{k \to 0} 
    {{\Gamma(K^-p \rightarrow \hbox{charged particles})} \over
     {\Gamma(K^-p \rightarrow \hbox{all final states})}} = 0.664 \pm 0.011 
                                                                   ,\\   
R_n & = & \displaystyle\lim_{k \to 0}
   {{\Gamma(K^-p \rightarrow \pi^0 \Lambda)} \over
    {\Gamma(K^-p \rightarrow \hbox{all neutral states})}} = 0.189 \pm 0.015 .
\end{eqnarray}
There are also data~\cite{Hem85} on the invariant mass spectrum of the 
$\Sigma^+ \pi^-$ system, which have been exploited to investigate the 
nature of the $\Lambda(1405)$ resonance.

Finally, the last piece of crucial experimental information comes from the 
recent KEK measurement~\cite{Iwa97} of the $K^-p$ scattering length.
As explained in the Introduction, the obtained value which includes the 
Coulomb effect:
$a^c_{K^-p} =      (-0.78 \pm 0.15  \pm 0.03) 
           + i   (0.49 \pm 0.25  \pm 0.12)\ \hbox{fm}$,
resolves the "kaonic hydrogen puzzle".

                       \subsubsection{\bf Results of the fit}
		       
\protect\label{subsec:resultfit}

For model OS1 interaction, we have adopted 
the same strategy as in Ref. \cite{Ose98}
by fitting our parameters $f$ and $\beta$ to the threshold branching ratios, 
with $f$ constrained to deviate from $f_{\pi}$ by less than $\pm 20\%$.
All other observables are "predicted", i.e. they are evaluated with
the values of the parameters reached at the end of minimization
(these values can be found in Ref. \cite{Baha02}).
The branching ratios and the $a_{K^-p}$ scattering length obtained in 
this model are comparable to the values from the Oset-Ramos model, 
see Table \ref{tab:bra}. 

\begin{table}[htb]
\squeezetable
\caption{$K^-p$ threshold strong branching ratios
and $K^-p$ scattering length (in fm), calculated in the particle basis. 
The results in the isospin basis are shown in italic letters.
See text for experimental data References.}
\protect\label{tab:bra}
\begin{center}
\begin{ruledtabular}
\begin{tabular}{lcccccccc}
Authors [Ref.] & $\gamma$ & $R_c$ & $R_n$ & & ${\cal{R}}e(a_{K^-p})$ & & 
                 ${\cal{I}}m(a_{K^-p})$ & Model  \\
\hline
Present work 
& 2.35 & 0.651 & 0.189 & & $-0.90$ & & 0.87 
& OSA \\

& {\it 3.17} &{\it 0.650} &{\it 0.257} & & ${\it-0.75}$ & & {\it1.11} 
& {\it isospin basis}\\

& 1.04 & 0.655 & 0.130 & & $-0.74$ & & 1.41 
& OSA, $\eta Y$ excluded \\
\\

& 2.36 & 0.657 & 0.193 & & $-0.98$ & & 0.80 
& OSB  \\
\\
Bahaoui et al.\cite{Baha02} 
& 2.38 & 0.636 & 0.171 & & $-1.04$ & & 0.83 
& OS1  \\
& {\it 3.37} & {\it 0.626} & {\it 0.244} & & ${\it -0.95}$ & &{\it  1.08} 
&  {\it isospin basis} \\
\\
Oset \& Ramos~\cite{Ose98} 
& 2.32 & 0.627 & 0.213 & & $-1.00$ & & 0.94 
& Chiral $\eta Y$ included \\

& {\it 3.29} & {\it 0.617} & {\it 0.292} & & ${\it -0.85}$ & & {\it 1.24} 
& {\it isospin basis} \\

& 1.04 & 0.637 & 0.158 & & $-0.68$ & & 1.64 
& Chiral $\eta Y$ excluded \\
\hline
Experiment 
& 2.36 $\pm$ 0.04 & 0.664 $\pm$ 0.011 & 0.189 $\pm$ 0.015 & & 
                  $-0.78 \pm 0.15 \pm 0.03$ & & 
		    0.49 $\pm$ 0.25  $\pm$ 0.12 &   \\
\end{tabular}
\end{ruledtabular}
\end{center}
\end{table}

\begin{figure}[htb]
\includegraphics[width=0.5\linewidth]{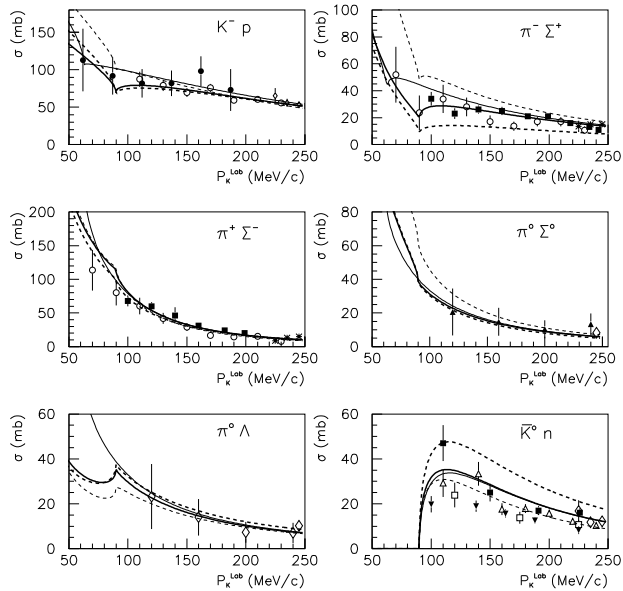} 
\caption{Total cross sections initiated by $K^-p$, calculated 
in the particle basis with different $\Kbar N$ models:
OSA (bold full line), OS1 (bold dashed line),
OSA-$\eta$ excluded (regular dotted line). The regular full line
is obtained with model OSA in the isospin basis.
Experimental data are from Refs. 
\cite{Abr65,Cse65,Sak65,Kit66,Kim67,Mast76,Bang81,Cibo82,Evans83}.}
\protect\label{fig:sigma}
\end{figure}

\begin{figure}[htb]
\includegraphics[width=0.5\linewidth]{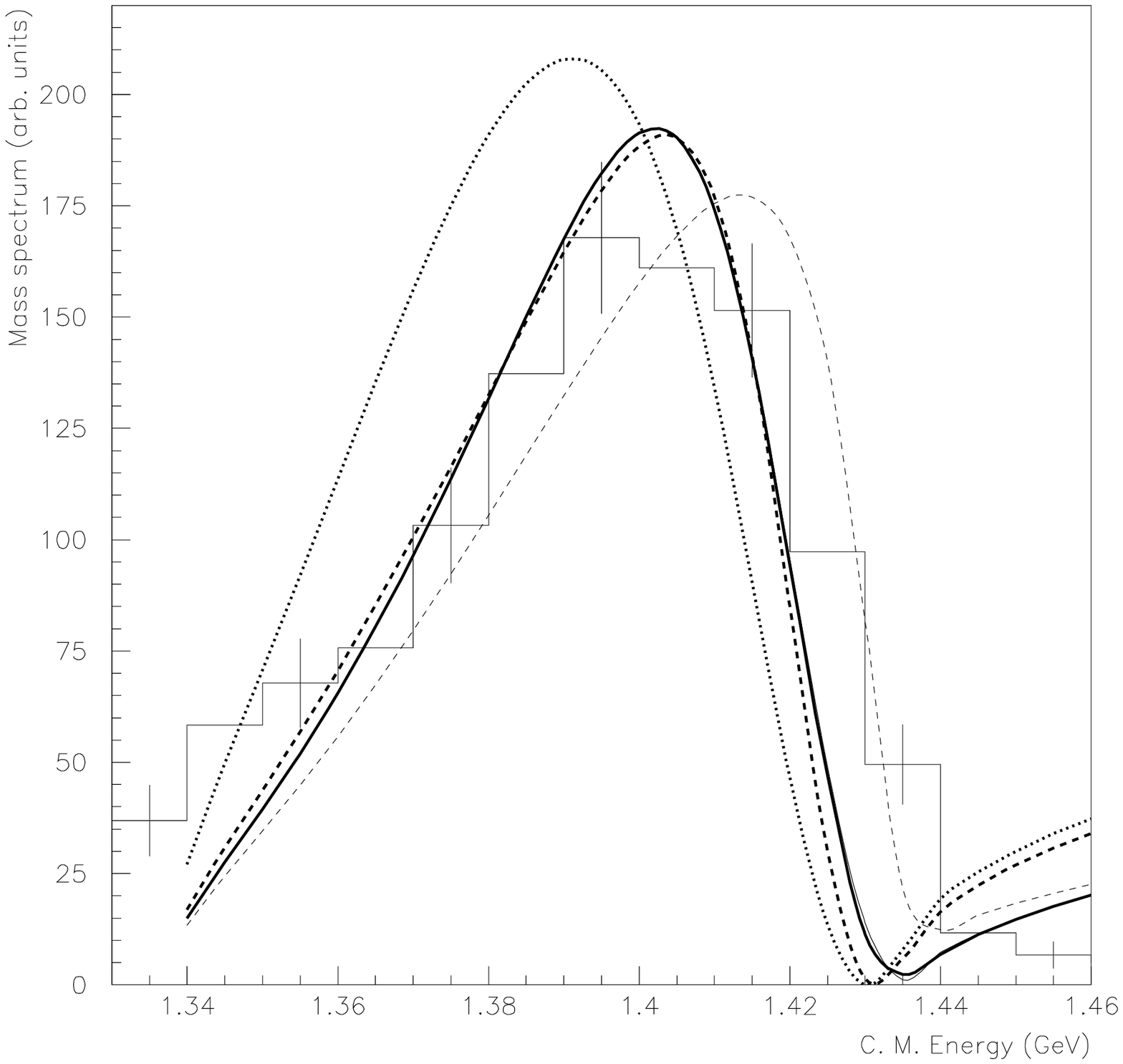} 
\caption{$\pi \Sigma$ mass spectrum, obtained with the same models
as in Fig. \ref{fig:sigma}. The bold dotted line is model OSB. 
The OSA results in the isospin and particle bases are practically identical. 
Experimental data are from Ref. \cite{Hem85}.}
\protect\label{fig:spectrum}
\end{figure}

The same conclusion holds when we
compare the total cross sections and the $\pi\Sigma$ mass spectrum given
in Figs. \ref{fig:sigma} and \ref{fig:spectrum} with the correponding results 
in Ref. \cite{Ose98}. 

For the OSA interaction, the parameters are
$f$, \ $\beta$, and $b^I_{ij}$'s, which are fitted to the 
threshold branching ratios and the $K^- p$ initiated 
cross sections, with the $b^I_{ij}$'s constrained to remain within $\pm 30\%$ 
of their exact SU(3)-symmetry values. The obtained values are given in 
Table \ref{tab:KN} 
\footnote{In our previous publication \cite{Baha02}, interaction OS2 
was introduced which
took an identical form to the current OSA.  The resulting fits by these 
two models are very similar.
The major difference is that the $SU(3)$ breaking has come out less in OSA
than in OS2, due to a more careful handling of the data-base.}  
..

\begin{table}[htb]
\squeezetable
\caption{Minimization results for the adjustable SU(3)-symmetry breaking 
coefficients of models OSA and OSB for the $\Kbar N$ interactions. 
For OS1, these coefficients are equal to 1, the average meson decay
constant is $f=1.20 f_{\pi}$, and the range is 870 MeV. 
We use $f_{\pi}$=93 MeV/c.
The reduced $\chi^{2}$ are roughly 1.2.}
\protect\label{tab:KN}
\begin{center}
\begin{ruledtabular}
\begin{tabular}{clcc}
Isospin & Channel & OSA & OSB \\
\hline
  & $f$      & 1.20 $f_{\pi}$ & 1.12 $f_{\pi}$ \\
\\
0 & $\Kbar N$                & 0.994 & 1.048  \\
  & $\Kbar N$-$\pi \Sigma$   & 1.108 & 0.917  \\
  & $\Kbar N$-$\eta \Lambda$ & 0.851 & 0.        \\
  & $\pi\Sigma$              & 0.903 & 0.926  \\
\\
1 & $\Kbar N$                & 1.056 & 0.833  \\
  & $\Kbar N$-$\pi \Sigma$   & 1.293 & 1.209  \\
  & $\Kbar N$-$\pi \Lambda$  & 0.943 & 0.933  \\
  & $\pi \Sigma$             & 0.991 & 1.290  \\
  & $\Kbar N$-$\eta \Sigma$  & 0.757 & 0.        \\
\\
  &  range (MeV)                   & 888.7 & 879.6  \\
\end{tabular}
\end{ruledtabular}
\end{center}
\end{table}

Note that in the lowest order chiral Lagrangian approach, some coupling 
coefficients: 
$C^{I=0}_{ij}=D_{ij},\ C^{I=1}_{ij}=F_{ij}$ as found in  Tables 2 and 3 
in Ref. \cite{Ose98} are equal to zero. We have chosen to keep these
zero values, thus we do not need the  
corresponding  $SU(3)$ breaking coefficients $b^I_{ij}$, which is reflected 
in Table \ref{tab:KN}.

Our results for OSA are presented in Table \ref{tab:bra} and 
Figs. \ref{fig:sigma}, \ref{fig:spectrum} along with available data 
as well as values from a few earlier models. 
We first discuss the result for $K^-p$ initiated  channels in the 
particle basis.
As shown in Table \ref{tab:bra},
the  threshold reaction ratios: $\gamma,\ R_c, \ R_n$  are better   
reproduced by OSA than by OS1. 
Regarding the $K^-p$ scattering length, our result is a prediction since 
we have not used the value extracted from the kaonic hydrogen 
atom \cite{Iwa97} as part of the constraint in the fitting procedure.  
The real part is obtained closer to the data \cite{Iwa97}  by OSA than 
by OS1. As for the imaginary part, both models give the values at the limit 
of the experimental uncertainties within 2 standard deviation. 
We point out that this trend for 
${\cal{I}}m(a_{K^-p})$ is systematically observed with either the present 
separable models or with the Oset-Ramos approach. The cross sections
are also better reproduced with OSA than with OS1, see Fig. \ref{fig:sigma}, 
especially those for the  $K^-p \to \pi^-\Sigma^+$ and 
the $K^-p \to \Kbar^\circ n$ 
channels which  now have the correct magnitude as compared with the data.
The position of the $\Lambda(1405)$ resonance predicted by the present models
are quite similar, and in good agreement with the data, 
see Fig. \ref{fig:spectrum} (and also Fig. \ref{fig:tkp}). 
Finally, we have calculated the cross section of the 
$K^- p \to \eta \Lambda$ reaction near threshold. We have found that 
the results predicted by models OSA and OS1 are in reasonable agreement 
with the recent data \cite{Sta01}
from the Crystal Ball detector at Brookhaven:
the steep rise juste above the threshold, and the $s$-wave behavior expected
for pseudoscalar meson production at threshold, are correctly reproduced.  

As for the  results obtained in the isospin basis we only discuss the case 
with the OSA interaction,
as the characteristic feature is the same with the result from the OS1 
interaction.
Here, we use the average masses of the hadron isospin multiplets everywhere
(see Table \ref{tab:masses}), hence now, for example, the charge exchange 
reaction: $K^-p \to \Kbar^{\circ} n$ becomes elastic energy-wise. 
Note that to calculate reaction cross sections, the entrance channel energies 
are taken using the physical particle masses, which are
also used to find the final state phase space volumes. 
In Table \ref{tab:bra}, we observe large differences between the values of 
the threshold ratios $\gamma$ and $R_n$ calculated in the two bases.
The shift of the cusps in the  cross sections
when using the isospin basis are clearly exhibited in Fig. \ref{fig:sigma}.
On the contrary, the position of the $\Lambda(1405)$ resonance is only slightly
affected. These differences
between the results in the two bases are similar to those observed  
in Ref.~\cite{Ose98}. 

Next we study the effect due to the contribution from the 
$\eta Y$ channels. Given that the difference between the two thresholds for 
$\eta \Lambda$ and
$K^-p$ final states is as large as $\sim 230$ MeV (with $\eta \Sigma$ the 
difference is about $75$ MeV larger), hence {\it a priori} the effect 
from the coupling  to these channels 
is expected to be insignificant. However, in Ref. \cite{Ose98} the authors 
observed the importance of
retaining these channels. So we want to check their claim. Very 
qualitatively, the role 
of these channels may be best understood  in the  exact isospin symmetry 
limit. Then 
$\eta Y$ channels are the only ones whose thresholds are above the one for 
the $\Kbar N$.
Hence, they provide a definite attraction to the elastic $\Kbar N$ 
process. As a result
the coupling to $\eta Y$ states controls the binding properties of the
$\Kbar N$ (in the effective chiral interaction adopted here, 
the $I=0$ $\Lambda (1405)$ is a bound state of $\Kbar N$ embedded 
in the continuum state of the $\pi \Sigma$ channel).  
As a first step, we have discarded these channels from
model OSA by forcing the $\Kbar N$-$\eta \Lambda$ and 
$\Kbar N$-$\eta \Sigma$ strengths to zero, without changing  other
parameters. Then, we have recalculated the amplitudes and observables.
This model will be hereafter referred to as "OSA, $\eta Y$ excluded".
The results are given in Table \ref{tab:bra} and
Fig. \ref{fig:sigma}. The only 
observables which are not affected are the threshold ratio $R_c$ and the 
$K^- p \to \pi^+ \Sigma^-$  cross section. All other quantities are
significantly modified, especially $\gamma$ and $R_n$ as well as the
$K^- p \to K^- p$ and $K^- p \to \pi^-\Sigma^+$ cross sections which
become unrealistic. Also, the maximum of the $\pi\Sigma$ mass spectrum 
is shifted towards the higher values of the momentum, thus incompatible 
with the data. Similar effects have been pointed out in Ref. \cite{Ose98}.
This situation can be understood by examining the values of the 
strength parameters in Table \ref{tab:KN}. The $\Kbar N$-$\eta Y$ strengths
deviate from unity by about $15\%$ to $25 \%$, and it is not possible
to obtain a correct overall fit if they are constrained to stay closer
to unity (in particular, the position of the $\Lambda(1405)$ resonance 
is not correctly reproduced).   
This does indicate that the $\Kbar N$-$\eta Y$ strengths take part in the 
minimization 
procedure at about the same level of importance as the other ones, and it is  
meaningless to turn them off entirely without re-adjusting other 
parameters. 
In order to see this problem from a somewhat  different angle 
we have introduced
yet another model: OSB, a variant of OSA in which the $\eta Y$ channels 
are excluded from the beginning.  
The values of the parameters are given in Table \ref{tab:KN}.
The results obtained for the $K^- p$ branching ratios and scattering length
are close to those obtained with OSA, see Table \ref{tab:bra}. This is also
the case for the total cross sections. However, the  position of
the $\Lambda(1405)$ resonance is now shifted towards  lower values 
of the momentum, see Fig. \ref{fig:spectrum} , which means that 
OSB is not able to reproduce 
the properties of the $\Lambda (1405)$. We thus need to retain the 
$\eta Y$ channels. 
  
Lastly, we use the parameters thus obtained both for OS1 and OSA to 
calculate the amplitudes (or $t$-matrices) for the $K^- n$ 
initiated processes, see Eq. (\ref{eq:ch1-5}),
for which, as mentioned earlier, there is no data to be confronted with.
We still need those amplitudes for our $K^-d$ three-body  calculations. 
Here along with the corresponding quantity in the $K^-p$ initiated channels,
we only present the scattering lengths as found in Table \ref{tab:AKN}.

\begin{table}[htb]
\squeezetable
\caption{$\Kbar N$ scattering lengths (in fm) calculated 
in the particle basis with models OS1 and OSA. The values in the last column
have been evaluated by Ramos \cite{Ram02} at the same energy.
$a_p$, $a_n^\circ$ and $a_{ex}$ (calculated at $W=M_{K^-}+M_p$) are 
the scattering lengths for elastic $K^-p$, $\Kbar^\circ n$, and 
charge exchange $K^-p \leftrightarrow  \Kbar^\circ n$, respectively. 
$a_n$ (calculated at $W=M_{K^-}+M_n$) is the scattering length 
for elastic $K^-n$.}
\protect\label{tab:AKN}

\begin{center}
\begin{ruledtabular}
\begin{tabular}{cccc}
Reactions & OSA & OS1  &    Oset-Ramos \\
\hline
$a_p(K^-p \to K^-p)$  & $ -0.888 + i\, 0.867 $ & $ -1.035 + i\, 0.828 $ 
                      & $ -1.013 + i\, 0.947 $ \\            
$a_n(K^-n \to K^-n)$  & $ \phantom{-} 0.544 + i\, 0.644 $ 
                                    & $ \phantom{-} 0.573 + i\, 0.452 $
		                    & $ \phantom{-} 0.540 + i\, 0.531 $ \\
$a_n^\circ(\Kbar^\circ n \to \Kbar^\circ n)$ 
                      & $ -0.444 + i\, 0.998 $ & $ -0.602 + i\, 0.894 $ 
                      & $ -0.516 + i\, 1.053 $ \\
$a_{ex}(K^-p \to \Kbar^\circ n)$  
                      & $ -1.215 + i\, 0.393 $ & $ -1.365 + i\, 0.484 $ 
                      & $ -1.289 + i\, 0.484 $ \\
\end{tabular}
\end{ruledtabular}
\end{center}
\end{table}
 
The differences between the results given by two models do not exceed 
$\sim 30 \%$. For comparison we give also the values obtained
by Oset and Ramos \cite{Ose98}. Our OS1 results agree with those values 
within $\sim 15 \%$, in line with the differences previously observed
on the other observables.  
The symmetry breaking effect in the mass of the hadron
isospin multiplets is quite visible, especially in the real parts,
since in the limit of isospin symmetry, one has: $a_p=a_n^\circ$,
and $a_{ex}=a_p-a_n$. Note that the scattering lengths in Table \ref{tab:AKN}
have been obtained at the $K^-p$ threshold (except for the elastic 
$K^-n$ process). 
In fact, these quantities are 
very sensitive to the value of the threshold at which they are calculated,
which is then reflected in the values obtained for the $A_{K^-d}$
scattering length. These aspects have been discussed in our 
previous paper \cite{Baha02}, and will be revisited at beginning
of Sec. \ref{sec:results}.

In conclusion, the OS1 and OSA parametrizations are good candidates to be used
in the three-body calculation. It is clear that such interactions,
the parameters of which are determined by a fit to the available observables 
in the particle basis with the chiral SU(3)-symmetry constraint, must 
definitely be
preferred  to previous separable interactions with
parameters determined in the isospin basis,
without any symmetry constraint, as it was the case
in Refs. \cite{Schick,Tok81,Tor86,Bah90,henleyetal}.


\subsection{$N N$ INTERACTION}
\protect\label{sec:nn_inter}

We have considered three different relativistic separable potentials to 
describe
the deuteron ($d$) channel. The structure of the equations may be read off
from the coupled-channels cases described in Appendix \ref{app-b}
by replacing the particle labels with the angular momentum labels 
relative to the $^3S_1$ and $^3D_1$ coupled partial waves. 

All the interactions considered are of rank-1. 
Characteristic to such potentials, the static parameters are 
correctly reproduced, namely: the triplet effective range parameters 
$a_t$ and $r_t$, the $D$-state percentage value $P_D$, the quadrupole 
moment $Q$, and the asymptotic ratio $\eta=A_D/A_S$. On the other hand, 
the $^3S_1$ and $^3D_1$ phase shifts cannot be reproduced simultaneously. 

The first model (hereafter called model A) is  the one proposed 
in our work on the $\pi d$ system \cite{Gir80}, namely the 
parametrization denoted by $SF(6.7)$ therein, with $P_D=6.7 \%$. This model is 
an extension of the usual Yamaguchi-type interactions, using form factors 
which are expressed as ratios of polynomials. The parameters are 
fitted to the static properties, the $^3S_1$ phase shift, and also to the 
deuteron monopole charge form factor up to about 6 fm$^{-1}$. 
All details can be found in Ref. \cite{Gir80}.

Besides this model  fitted to on-shell properties only,
we have considered a second model derived from the PEST1 potential
constructed by Haidenbauer and Plessas in Ref. \cite{Hai84}.
The authors have constructed 
a separable representation of the Paris potential to reproduce
both its on-shell and off-shell characteristics. Among  various $NN$
partial waves, special care is devoted to the coupled $^3S_1$-$^3D_1$ state.
The best approximation to the Paris potential requires a rank-4
interaction. However, for applications where only the deuteron bound-state
enters, a rank-1 parametrization was proposed, called PEST1, with 
all deuteron properties (including the wave functions) being practically 
the same as those given by the Paris potential. The price to pay is 
that the form
factors are chosen as sum of rational functions, with many parameters, 
see Ref. \cite{Hai84}. 
As the PEST1 parametrization is non-relativistic, we have extended it 
by means of {\it the minimal relativity rule}. More specifically,  
the relativistic potential $V^R$ between nucleons 1 and 2 is obtained 
from the non-relativistic one $V^{NR}$ according to the following
transformation in momentum space:

\begin{equation}
   V^R(p,p') = (2\pi)^3 \sqrt{2 \epsilon_{1p} 2 \epsilon_{2p}} V^{NR}(p,p')
                        \sqrt{2 \epsilon_{1p'} 2 \epsilon_{2p'}} ,
\end{equation}

\noindent
with $\epsilon_{ip} = \sqrt{{\bm p}_i^{\,2} + m_i^2}$
the total energy of nucleon $i$.
Taking $V^{NR}$ as separable:  $V^{NR}(p,p')=\lambda g(p) g(p')$,
$V^{R}(p,p')$ has the same form, with $g(p)$ multiplied by:
$2  \sqrt{(2\pi)^3} \sqrt{{\bm p}^{\,2} + m^2}$.
We have checked that this transformation induces only slight changes
in the deuteron properties. For example, the original $P_D$ value is $5.8\%$, 
and the value obtained after the relativistic transformation is $6.1\%$. 
So, we can keep the original parameters.
In our present work this is denoted as model B. 
 
Finally, in order to assess the importance of the $D$-state contribution
in the low energy $K^- d$ scattering, we have used a pure $^3S_1$ relativistic
potential, with form factor as given in Eq. (\ref{eq:gp}). The values of the 
strength and range parameters fitted to $E_D$ and $a_t$ are:
$\lambda_d=-5974.2$,  $\beta_d=1.412$ fm$^{-1}$, respectively. 
This model is called  model C.


\subsection{OTHER TWO-BODY INPUT}
\protect\label{sec:other_input}

We have also considered the contributions of the $\pi N$-$P_{33}$ 
and coupled $Y N$ interactions. 
As will be explained in Sec. \ref{sec:results_c}, these two-body channels 
start to contribute from the second order in the multiple scattering 
expansion of the Faddeev equations, so 
their contribution in the low energy $K^- d$ observables is expected to be 
moderate or even small.

         \subsubsection{\bf $\pi N$-$P_{33}$ interaction}
 
We have chosen a conventional $\Delta$-isobar model where the 
$\Delta$ resonance
is parametrized according to a one-term separable potential:

\begin{equation}
V(p , p' ; \sigma) =  g(p) \frac{\lambda_{\Delta}}{\sigma-m_0^2} g(p') .
\label{eq:vp33}
\end{equation} 

\noindent
Here, $\sigma$ is the $\pi N$ total c.m. energy squared, and $m_0$ the bare 
$\Delta$ mass. A $p$-wave monopole form factor is assumed with 
cut-off $\Lambda$:

\begin{equation} 
g(p)= \frac{p}{p^2+\Lambda^2} . 
\label{eq:gp33}
\end{equation}

The strength parameter $\lambda_{\Delta}$, together with $m_0$ and $\Lambda$,
are fitted to the $P_{33}$ phase-shift. In the relativistic approach, we have 
obtained the following values: $\lambda_{\Delta}=218.582$, $m_0=1308.8$ MeV,
and $\Lambda=290.9$ MeV/c. 

In order to solve the three-body equations in the particle basis, it is 
necessary to introduce different charge states of the pion and nucleon.
So, the following states will contribute: the $(\pi^0 p, \pi^+ n)$ and  
$(\pi^- p, \pi^0 n)$ coupled states, and the $(\pi^- n)$ state. 
The corresponding separable two-body potentials are obtained by choosing
the same form factors as Eq. (\ref{eq:gp33}) in all channels, and expressing
the strength parameters in terms of $\lambda_{\Delta}$, 
see Appendix~\ref{app-a}. The resulting two-body $t$-matrices entering  
the three-body equations are obtained straightforwardly.

	 \subsubsection{\bf $Y N$ interactions}

The hyperon-nucleon interaction has been well studied in the past.
One of the most popular approaches is  meson-exchange potential models 
with $SU(3)$
symmetry constraints  used in the coupled channels equations, see Ref. 
\cite{Rij99} and references therein. 
Besides, the intrinsic interest in investigating the available 
$YN$ experimental data, these interactions serve as input to  
hypernuclear physics.
Concurrently, separable models have been developed, and some of them have
been used as input to $K^-d$ three-body calculations. In particular, 
the effect of the final state $YN$ interaction (limited to $s$-waves) on
the $\Lambda p$ invariant mass distribution near the 
$\Sigma N$ threshold was studied in Refs. \cite{Tok81,Tor86}. 
One of the main conclusion was that the best reproduction of the shoulder 
in the $\Lambda p$ mass spectrum favoured  models which do not support 
an unstable $\Sigma N$ bound state. The interactions that we have elaborated
in Ref. \cite{Bah90,Bah_thesis} and the one presented in this work fulfill 
this condition.

The data to which the adjustable parameters of the separable potential
should  be fitted are scarce and exhibit rather large error bars.
To our knowledge, there are no new data in addition to those used in
our previous work, namely:
(i) the $\Sigma^+p \to \Sigma^+p$, $\Sigma^-p \to \Sigma^-p$,
$\Sigma^\circ n$, $\Lambda n$, and  $\Lambda p \to \Lambda p$ 
total cross sections, 
in the hyperon momentum range $p^{lab}_Y \le 300$ MeV/c,
(ii) the $\Sigma^+p \to \Sigma^+p$, $\Sigma^-p \to \Sigma^-p$, and  
$\Sigma^-p \to \Lambda n$ differential cross sections 
for $p^{lab}_Y \sim  300$ MeV/c. See Refs. 
\cite{Alex68,Sech68,Eis71,Eng66,Hep68,Kad71,Hau77}. 

In Refs. \cite{Bah90,Bah_thesis}, we have used most of these data to
determine the parameters of two models: one non-relativistic, and the 
other relativistic. The calculations were done in the isospin basis,
where one must consider the $\Sigma N$-$\Lambda N$ ($I=1/2$) 
coupled channels and the $\Sigma N$ ($I=3/2$) single channel. 
The main conclusion
was that the total cross sections are dominated by the $^3S_1$ partial wave, 
except for $\Sigma^+p \to \Sigma^+p$ which is dominated by the $^1S_0$ wave, 
while the $P$ and $D$ partial waves have a significant contribution only in 
the $\Sigma^-p \to \Lambda n$ total cross section (of course, these 
higher partial waves must be taken into account if the differential cross 
sections are added to the data to be fitted).
 
\begin{figure}[htb]
\includegraphics[width=0.5\linewidth]{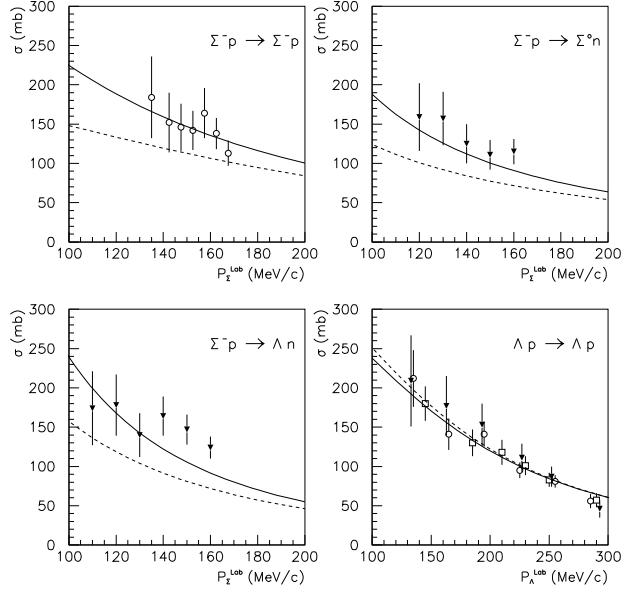} 
\caption{$\Sigma^-p \to \Sigma^-p$, $\Sigma^0n$, $\Lambda n$, and  
$\Lambda p \to \Lambda p$ total cross sections, calculated in the 
particle basis (full line) and in the isospin basis (dashed line).
Experimental data are from Refs.
\cite{Alex68,Sech68,Eis71,Eng66,Hep68,Kad71,Hau77}.}
\protect\label{fig:snln}
\end{figure}

The model used in the present study is an extension of the above mentioned
relativistic model to calculate the observables in the particle basis.
Only the $^3S_1$ partial wave contributions are included, thus the 
parameters are fitted to the total cross sections, except for 
$\Sigma^+p \to \Sigma^+p$. Note that neglecting the $^1S_0$ partial wave
is justified in the three-body calculation at low energies, since the 
contribution from the singlet $S$-wave $YN$ interactions 
is excluded for parity considerations.
We take as adjustable parameters the coupling strengths and the ranges 
of the form factors in the isospin basis. The observables are calculated 
in the particle basis, where the following channels contribute: 
the $(\Sigma^\circ p, \Sigma^+ n, \Lambda p)$ and 
$(\Sigma^- p, \Sigma^\circ n, \Lambda n)$ coupled channels, and the
$\Sigma^- n$ and  $\Sigma^+ p$ single channels. The relations between
the strengths parameters in the two bases can be found in 
Appendix \ref{app-a},  
and the transition matrices for the different reactions are obtained 
from the general expressions given in Appendix \ref{app-b}. 
We take monopole form factors, Eq. (\ref{eq:gp}), with isospin-independent
ranges. 
The values of the fitted parameters are given in Table~\ref{tab:SNLN},
and the cross sections in Fig. \ref{fig:snln}. The selected data can be
well reproduced by this simple model, but a large reduction effect
is observed when isospin basis is adopted in the total cross sections
for the $\Sigma^- p$ induced reactions at low values of the hyperon momentum.

We give in Table \ref{tab:ldif-snln} the $YN$ scattering lengths calculated
in the particle basis. 

\begin{table}[htb]
\squeezetable
\caption{Strength and range parameters of the $Y N$
interactions.}
\protect\label{tab:SNLN}

\begin{center}
\begin{ruledtabular}
\begin{tabular}{clc}
Isospin & Channel &  Parameters \\
\hline
1/2 & $\Sigma N$                & -3209.69  \\
  & $\Sigma N$-$\Lambda N$      & -1739.22  \\
  & $\Lambda N$                 & -1794.16  \\
3/2 & $\Sigma N$                &  3072.26  \\
  & $\Sigma N$ range (MeV)          & 452.300  \\
  & $\Lambda N$ range (MeV)          & 356.981  \\  
\end{tabular}
\end{ruledtabular}
\end{center}
\end{table}

Except for $a_{\Sigma^+ p - \Sigma^+ p}$ and 
$a_{\Sigma^- n - \Sigma^- n}$ which are practically equal, the symmetry 
breaking effects are large. For example, at the limit of exact isospin
symmetry, one should have: 
$a_{\Sigma^- p - \Sigma^\circ n} = 
\sqrt 2 (a_{\Sigma^\circ n - \Sigma^\circ n}-a_{\Sigma^- p - \Sigma^- p})=
-a_{\Sigma^+ n - \Sigma^ p}=
\sqrt 2 (a_{\Sigma^\circ p - \Sigma^\circ p}-a_{\Sigma^+ n - \Sigma^+ n})$,
which is clearly not the case.

\begin{table}[htb]
\squeezetable
\caption{$YN$ scattering lengths (in fm) calculated at $W=M_{\Sigma^+}+M_n$
for channels $(\Sigma^+ n, \Sigma^\circ p, \Lambda p)$, 
and $W=M_{\Sigma^-}+M_p$
for channels $(\Sigma^- p, \Sigma^\circ n, \Lambda n)$. 
The single channels values are :
$a_{\Sigma^- n} = -0.454 $ ($W=M_{\Sigma^-}+M_n$), and 
$a_{\Sigma^+ p} = -0.455 $ ($W=M_{\Sigma^+}+M_p$).}
\protect\label{tab:ldif-snln}

\begin{center}
\begin{ruledtabular}
\begin{tabular}{cccccccccc}
    channel & $\Sigma^+ n$ & $\Sigma^0 p$ & $\Lambda p$ & channel&
              $\Sigma^- p$ & $\Sigma^0 n$ & $\Lambda n$ & \\
\hline
       $\Sigma^+ n$
     & $ 0.609 + i\, 3.618 $ & $ 0.834 + i\, 2.837$ & $-0.113 - i\, 1.595$
     & $\Sigma^- p$
     & $ -0.528 + i\, 2.505$ & $-0.192 - i\, 1.762$ & $0.241 - i\, 0.996$ \\
     $\Sigma^0 p$ 
     &
     & $0.145 + i\, 2.225$ & $-0.089 - i\, 1.251$ 
     & $\Sigma^0 n$
     & 
     & $-0.137 + i\, 1.277$ & $-0.072 + i\, 0.721$ \\
     $\Lambda p$ 
     &   
     & 
     & $1.893$
     & $\Lambda n$
     & 
     & 
     & $1.851$ \\
\end{tabular}
\end{ruledtabular}
\end{center}
\end{table}


\section{Three-body equations for the $K^- d$ system}
\protect\label{sec:theory}
%
In this Section, we describe the three-body equations for the
$K^- d$ system, in which the two-body input described in the 
previous Section will enter.   
In the first Subsection, we recall the general form of the three-body 
equations written in the isospin basis, in the case of coupled two-body 
input channels, and we give the equations for the rotationally invariant 
amplitudes. 
Then, the antisymmetrization due to the identity of the two mucleons in the
isospin basis is examined. In the second Subsection, the extension 
to the particle basis is given. Some important aspects concerning the 
practical calculation are considered in the third Subsection.

         \subsection{Three-body equations in the isospin basis}

The extension of the usual three-body equations to the case where the two-body 
operators connect two states involving particles which are different
(i.e. inelastic channels), 
results in the following system of coupled equations, written in operator 
form:

\begin{equation}
\label{eq:3body}
X_{a\,  b}(s) = Z_{a\, b}(s) + \sum_{c,\,c'} Z_{a\, c}(s) 
                                  R_{c\, c'}(s)\; X_{c'\, b}(s) ,
\end{equation}  

\noindent
with $s$ the 3-body total energy.
Here, $a,b,c$ and $c'$ are the indices which specify the particles
involved in the various three-body channels, namely the spectator and the
interacting pair. $X_{ab}$ is the transition amplitude between channels
$a$ and $b$, and  $Z_{ab}$ is the corresponding Born term.
The main difference with respect to the usual case is that the two-body
operator $R_{c\, c'}$ connects two different two-body states labeled
as $c$ and $c'$. 

We now specify the values taken by the channel indices 
in Eq. (\ref{eq:3body}).
Taking into account all the two-body input considered in 
Section \ref{sec:two-body}, one must consider the following types of 
three-body channels in the isospin basis: $d\,[\,\Kbar(NN)\,]$, 
$y\,[\, N(\Kbar N)\,]$, $\alpha\,[\,N(\pi Y)\,]$, $\mu\,[\,N(\eta Y)\,]$,
$\,\beta\,[\,\pi (YN)\,]$, and $\Delta\,[\,Y(\pi N)\,]$. 
Here, the first letter is the {\it label} of the channel,
and in square braces we specify the {\it spectator} and the 
associated {\it pair} of particles. In fact, labels $y$, $\alpha$, $\mu$
and $\beta$ can be considered as "generic" names. In practice, extra indices
are needed to fully describe the physical situation. For example,
for the pairs corresponding to the $\Kbar N$ interactions, there are 
three $I=0$ coupled channels ($\Kbar N$-$\pi \Sigma$-$\eta \Lambda)$, 
and four $I=1$ coupled channels 
($\Kbar N$-$\pi \Sigma$-$\pi \Lambda$-$\eta \Lambda)$). 
A similar situation occurs in the case of the $\beta$ channels corresponding 
to the $\Sigma N$-$\Lambda N$ ($I=1/2$)  coupled channels. The corresponding
three-body channels are labeled as summarized in Table \ref{tab:labels}.
Note that in channel $\Delta$, the spectator hyperon $Y$ is restricted 
to $\Sigma$, from isospin considerations.

\begin{table}[htb]
\squeezetable
\caption{Labels of the three-body channels. The second line specifies the
isospin of the two-body sub-system.}
\protect\label{tab:labels}
\begin{center}
\begin{ruledtabular}
\begin{tabular}{ccccccccccccc}
  $\mbox{channel}$  & $\Kbar (NN)$ &  
                      $N(\Kbar N)$ & $N(\pi\Sigma)$ & $N(\eta\Lambda)$ &
                      $N(\Kbar N)$ & $N(\pi\Sigma)$ & $N(\pi\Lambda)$  &
		                                      $N(\eta\Sigma)$  &
		      $\Sigma(\pi N)$   & 
                      $\pi(\Sigma N)$   & $\pi(\Lambda N)$ &
		      $\pi(\Sigma N))$\\
$\mbox{isospin}$ & 0 & 0 & 0 & 0 & 1 & 1 & 1 & 1 & 3/2 & 1/2 & 1/2 & 3/2 \\
\hline
 label & $d$      & 
         $y_1$    & $\alpha_1$ & $\mu_1$    &
	 $y_2$    & $\alpha_2$ & $\alpha_3$ & $\mu_2$    &
	 $\Delta$ & $\beta_1$  & $\beta_2$  &
	 $\beta_3$  \\
\end{tabular}
\end{ruledtabular}
\end{center}
\end{table}

In the formal equations (\ref{eq:3body}), the channel indices $a,b,c,c'$ 
take their values in the set defined above: $\{d,y,\alpha,\mu,\beta,\Delta\}$, 
and the different quantities: $X$, $Z$, $R$, can be considered as 
matrices with respect to these indices.

The transitions between different two-particle states are induced by 
the two-body operators $R_{c\, c'}$: 
for example, $R_{y\alpha}$ is the quantity $R_{\Kbar N-\pi Y}$ corresponding
to the two-body $t$-matrix for the $\Kbar N \to \pi Y$ transition, 
evaluated in the presence of a spectator nucleon.  
Written in matrix form, the non-zero $R$ operators appear as 
block-matrices as shown in Table~\ref{tab:propags}.

\begin{table}[htb]
\squeezetable
\caption{Matrix of propagators in the isospin basis.}
\protect\label{tab:propags}
\begin{center}
\begin{ruledtabular}
\begin{tabular}{cccccccccccccc}
 channel & & $d$      & 
             $y_1$    & $\alpha_1$ & $\mu_1$ &
	     $y_2$    & $\alpha_2$ & $\alpha_3$ & $\mu_2$ &
	     $\Delta$ & $\beta_1$  & $\beta_2$  &
	     $\beta_3$ \\
\hline
$\Kbar (NN)$  & $d$ & $R$  &  &  &  &  &  &  &  &  &  &  & \\
\hline
$N(\Kbar N)$  & $y_1$ & & $R$ & $R$ & $R$ &  &  &  &  &  &  &  &  \\
$N(\pi\Sigma)$ & $\alpha_1$ &  & $R$ & $R$ & $R$ &  &  &  & &  &  &  & \\
$N(\eta\Lambda)$ & $\mu_1$ &  & $R$ & $R$ & $R$ &  &  &  & &  &  &  & \\
\hline
$N(\Kbar N)$  & $y_2$ & & & & & $R$ & $R$ & $R$ & $R$ &  &  &  &  \\
$N(\pi\Sigma)$ & $\alpha_2$ & & & & & $R$ & $R$ & $R$ & $R$ &  &  &  &  \\
$N(\pi\Lambda)$ & $\alpha_3$ & & & & & $R$ & $R$ & $R$ & $R$ &  &  &  &  \\
$N(\eta\Sigma)$ & $\mu_2$ & & & & & $R$ & $R$ & $R$ & $R$ &  &  &  & \\
\hline
$\Sigma(\pi N)$ & $\Delta$ & & & & & & & & & $R$ &  &  & \\
\hline
$\pi(\Sigma N)$ & $\beta_1$ & & & & & & & & & & $R$ & $R$  &  \\
$\pi(\Lambda N)$ & $\beta_2$ & & & & & & & & & & $R$ & $R$  & \\
\hline
$\pi(\Sigma N)$ & $\beta_3$ & & & & & & & & & & & & $R$  \\
\end{tabular}
\end{ruledtabular}
\end{center}
\end{table}
 
Concerning the Born terms, only those which connect initial and final 
states involving the same three particles are different from zero. 
These terms are shown in matrix form in Table \ref{tab:born-terms}.

\begin{table}[htb]
\squeezetable
\caption{Matrix of Born terms in the isospin basis. 
For the non-zero Born terms, the exchanged particle is shown in parentheses.}
\protect\label{tab:born-terms}
\begin{center}
\begin{ruledtabular}
\begin{tabular}{cccccccccccccc}
 channel & & $d$      & 
             $y_1$    & $\alpha_1$ & $\mu_1$ &
	     $y_2$    & $\alpha_2$ & $\alpha_3$ & $\mu_2$ &
	     $\Delta$ & $\beta_1$  & $\beta_2$  &
	     $\beta_3$ \\
\hline
$\Kbar (NN)$  & $d$ &   & $Z(N)$ &  &  & $Z(N)$ &  &  &  &  &  &  & \\
\hline
$N(\Kbar N)$  & $y_1$ & $Z(N)$ & $Z(\Kbar)$ & & & $Z(\Kbar)$ & & & & & & & \\
$N(\pi\Sigma)$ & $\alpha_1$ & & & & & & & & & $Z(\pi)$ & $Z(\Sigma)$ & & 
$Z(\Sigma)$ \\
$N(\eta\Lambda)$ & $\mu_1$  &  &  &  &  &  &  &  &  &  &  &  &  \\
\hline
$N(\Kbar N)$  & $y_2$ & $Z(N)$ & $Z(\Kbar)$ & & & $Z(\Kbar)$ & & & & & & & \\
$N(\pi\Sigma)$ & $\alpha_2$ & & & & & & & & & $Z(\pi)$ & $Z(\Sigma)$ & & 
$Z(\Sigma)$ \\
$N(\pi\Lambda)$ & $\alpha_3$ & & & & & & & & & & & $Z(\Lambda)$ & \\
$N(\eta\Sigma)$ & $\mu_2$  &  &  &  &  &  &  &  &  &  &  &  & \\
\hline
$\Sigma(\pi N)$ & $\Delta$ & & & $Z(\pi)$ & & & $Z(\pi)$ & & & & $Z(N)$ 
& & $Z(N)$ \\
\hline
$\pi(\Sigma N)$ & $\beta_1$ & & & $Z(\Sigma)$ & & & $Z(\Sigma)$ & & & $Z(N)$ 
& & \\
$\pi(\Lambda N)$ & $\beta_2$ & & & & & & & $Z(\Lambda)$ & & & & & \\
\hline
$\pi(\Sigma N)$ & $\beta_3$ & & & $Z(\Sigma)$ & & & $Z(\Sigma)$ & & & $Z(N)$ 
&  &  & \\
\end{tabular}
\end{ruledtabular}
\end{center}
\end{table}
 
For example, $Z_{dy}$ is the Born term for the exchange of a nucleon 
between the $\Kbar(NN)$ and $N(\Kbar N)$ states. Note that $Z_{dd}$ is 
equal to zero, since no particle can be exchanged between the initial 
and final $NN$ pairs. We note also that, using the two-body input channels
considered here, there is no Born term involving an $\eta Y$ pair. 
To have such terms, it would be neccessary to take into account the
contributions of three-body channels like $\eta(YN)$, $Y(\pi N)$
and $Y(\eta N)$. (the last two channels necessitate to introduce
the $\pi N$-$\eta N$ two-body coupled system). These contributions are
expected to be negligible at the $K^- d$ threshold.  

The successive steps leading from the formal equations (\ref{eq:3body})
to the relativistic equations for the rotationally invariant amplitudes 
are the same as in the usual case. We refer the reader to References 
\cite{Bah_thesis,Aar77,Gir78,Aar68,Rin77} for all details. 
The final equations read:

\begin{equation}
\label{eq:rotinv}
X_{\tau_a \tau_c}^{\cal JI}(p_a , p_c ; s) =
   Z_{\tau_a \tau_c}^{\cal JI}(p_a , p_c ; s) + \sum_{b,\tau_b ; b',\tau_{b'}}
   \int\frac{dp_b \, p_b^2}{2\epsilon_{b}}
   Z_{\tau_a \tau_b}^{\cal JI}(p_a , p_b ; s)R_{bb'}^{c_b=c_{b'}}(\sigma_{b})
   X_{\tau_{b'} \tau_c}^{\cal JI}(p_b , p_c ; s) , 
\end{equation} 

\noindent
where  $\sigma_{b}$ is the invariant energy of the pair in channel $b$
expressed in the three-body center of mass system,
$c_a = (J_a, S_a, I_a)$ specifies the conserved quantum numbers of the
pair in channel $a$, and $\tau_a=(c_a, l_a,\Sigma_a)$ specifies the 
three-body quantum numbers in this channel. Note that labels $c$ and $\tau$
refer to the spin-isopin variables in a given channel. For example,
assuming that channel $a$ is composed with particle $i$ as spectator
and the pair $(jk)$, we define the following quantities:

-- $\bm s_i$ : spin of particle $i$ ,

-- $\bm S_i$ $(= \bm s_j + \bm s_k)$, $\bm L_i$, and 
   $\bm J_i$ $(= \bm L_i + \bm S_i)$ : spin, orbital angular momentum,
and total angular momentum, respectively, of pair $(jk)$ ,

-- $\bm \Sigma_i$ $( = \bm s_i + \bm J_i)$, $\bm l_i$, and
   $ \cal J$ $(= \bm l_i + \bm \Sigma_i)$ :  channel spin, 
   orbital angular momentum of $i$ and $(jk)$, and  three-body 
   total angular momentum, respectively.

\noindent
The isospin variables are defined in the same way:

-- ${\bm t_i} $ : isospin of particle $i$ ,

-- $\bm I_i$ $( = \bm t_j + \bm t_k)$ : isospin of pair $(jk)$,

-- $ \cal I$ $( = \bm t_i + \bm I_i)$ : three-body total isospin.

The notations used in Eq. (\ref{eq:rotinv}) for the two-body propagators 
are the following:
the lower indices $b$ and $b'$ refer to the involved coupled channels,
and the upper "index" $c_b=c_{b'}$ means that the spin-isospin 
quantum numbers are conserved by the interaction, i.e.:  
$J_b=J_{b'}$, $S_b=S_{b'}$, $I_b=I_{b'}$ (the latter equality holds only
in the isospin basis). For the uncoupled propagators,
the single index $c_b$ specifies completely the interacting pair. 

The two-body propagators are calculated as explained in 
Section \ref{sec:two-body} and Appendix \ref{app-b}, and the general 
expression of the Born term can be found in Refs. \cite{Bah_thesis}.

We end this Subsection with two specific aspects concerning the
calculation of the Born terms in the isospin basis. The first one
concerns the problem of antisymmetrization.
In the isospin basis, the particles in the different multiplets 
are considered as identical. In particular, the neutron and proton 
are treated as identical particles: the nucleon $N$.
Therefore, one must properly antisymmetrize the amplitudes and Born terms 
where the initial and final three-body states 
involve two nucleons. As a result, one must introduce antisymmetrization 
coefficients as explained in Appendix \ref{app:antisym}.
The second aspect concerns the problem relative to the "ordering" of
particles when evaluating the Born terms. For example, let us
consider the $Z_{yy}$ Born term for $\Kbar$ exchange between two 
$\Kbar N$ states, with a spectator nucleon. The antisymmetrized expression 
is, according to Appendix \ref{app:antisym}:

\begin{equation}
\label{eq:zyy_sym}
\widetilde{Z}_{yy} = - Z_{y^1y^2} =
                     - \bra{N_2(\Kbar N_1)} G_0 \ket{N_1(\Kbar N_2)} , 
\end{equation}

\noindent
with $G_0$ the three-body propagator.
Usually, the expressions given for the Born terms assume a cyclic ordering
of the particle labels, see for example Refs. \cite{Bah_thesis,Fay71}. 
Namely, one calculates for example: $Z_{ij}=\bra{i(jk)} G_0 \ket{j(ki)}$, 
for the exchange of particle $k$
between pairs $(jk)$ and $(ki)$, where $i$, $j$, $k$ are assumed to be
cyclically ordered both in the initial and final states. In the case of 
$Z_{y^1y^2}$, Eq. (\ref{eq:zyy_sym}), the spectator particles in the
initial and final states are labeled as: $N_2=i$, $N_1=j$, thus the
Born term has the "non-cyclic" form:  $\bra{i(kj)} G_0 \ket{j(ki)}$.
We obtain the cyclic form by exchanging particles $j$ and $k$ 
(i.e. $N$ and $\Kbar$), in the pair of the final state.
This introduces the following phase coefficient:

\begin{equation}
(-1)^{L_{\Kbar N}} (-1)^{S_{\Kbar N}-s_N-s_{\Kbar}}
                   (-1)^{I_{\Kbar N}-t_N-t_{\Kbar}} .
\end{equation} 

\noindent
The first factor is due to changing the direction of the relative momentum
of the pair, and the second (third) factor results from the property
of the Clebsch-Gordan coefficients when the coupling order of the spins 
(isospins) of two particles is changed.

%
             \subsection{\bf Extension to particle basis}

In the particle basis, we consider the deuteron as composed of two distinct
particles: the neutron ($n$) and the proton ($p$), and the particles 
of the different multiplets take their physical masses. The number of 
three-body channels to be considered 
increases considerably as compared to the isospin basis case. For example,
we have seen in Section \ref{sec:kbarn_inter} that for the $\Kbar N$ 
interactions in the particle basis, we must consider the coupled 
channels related to $K^-p$, namely: 
$K^-p$, $\Kbar^\circ n$, $\pi^-\Sigma^+$, $\pi^+\Sigma^-$, 
$\pi^\circ\Sigma^\circ$, 
$\pi^\circ\Lambda$, $\eta\Sigma^\circ$, $\eta\Lambda$, 
and those related to $K^- n$: 
$K^- n$, $\pi^-\Sigma^\circ$, $\pi^\circ\Sigma^-$, $\pi^-\Lambda$, 
$\eta\Sigma^-$.
So, if we retain only these contributions in addition to the deuteron, 
the three-body channels to be considered are: the $K^-(pn)$ channel, 
the eight above channels with the neutron as spectator, and the five 
remaining channels with the proton as spectator. So, we have a 14$\times$14
Born terms matrix. However, the number of non-zero Born terms is very limited.
In the case considered here, we have only the following different Born terms:
$\bra{K^-(pn)} G_0 \ket{n(K^-p)}$ (proton exchange between the deuteron 
and the $(K^-p)$ pair), 
$\bra{K^-(pn)} G_0 \ket{p(K^-n)}$ (neutron exchange between the deuteron 
and the $(K^-n)$ pair),
$\bra{n(K^-p)} G_0 \ket{p(K^-n)}$ ($K^-$ exchange between the $(K^-p)$ 
and $(K^-n)$ pairs),
$\bra{n(\Kbar^\circ n)} G_0 \ket{n(\Kbar^\circ n)}$ ($\Kbar^\circ$ exchange 
between the initial and final $(\Kbar^\circ n)$ pairs),
and the symmetric terms. 

The matrix of propagators has a block structure similar to that
in the isospin basis (Table~\ref{tab:propags}~): 
single term for the deuteron propagator, 8$\times$8 matrix for the 
channels coupled to $K^-p$,
and 5$\times$5 for the channels coupled to $K^-n$.

If we take into account the contributions of the $\pi N$ and $Y N$ 
interactions, the following additionnal three-body channels must be 
considered (see Section \ref{sec:other_input}): 
$\{\Sigma^-(\pi^\circ p), \Sigma^-(\pi^+ n)\}$, 
$\{\Sigma^\circ(\pi^- p), \Sigma^\circ(\pi^\circ n)\}$, and        
$\Sigma^+(\pi^- n)$, 
coming from the $\pi N$ interaction,
and: 
$\{\pi^-(\Sigma^+ n), \pi^-(\Sigma^\circ p), \pi^-(\Lambda p)\}$,
$\{\pi^\circ(\Sigma^- p), \pi^\circ (\Sigma^\circ n), 
\pi^\circ(\Lambda n)\}$, and
$\pi^+(\Sigma^- n)$,
coming from the $Y N$ interactions. 

The derivation of the equations for the rotationally invariant amplitudes, 
and the calculation of the Born terms, are done along the same lines as in 
the isospin basis. The main difference is that there is no problem
relative to antisymmetrization, except for 
$\bra{n(\Kbar^\circ n)} G_0 \ket{n(\Kbar^\circ n)}$ which must be
antisymmetrized with respect to the two identical neutrons, as
in the case of the $Z_{yy}$ Born term calculated in the isospin basis,
see Appendix \ref{app:antisym}.
Another important difference concerns the 
isospin dependence. In the isospin basis,
the expression of the Born term involves a "6-j" coefficient
originating from the transformation from  initial to final 
three-particle isospin states, see Ref. \cite{Bah_thesis,Fay71}. 
This coefficient depends on the values of the particle isospins, 
total isospin of the initial and final pairs,
and three-body total isospin. In the particle basis, the individual
isospins of all particles are well defined, but not the total isospin 
of the pairs (consequently, the $c$ labels in Eq. (\ref{eq:rotinv})
do not depend on the isospin $I$ of the pairs). 
As the initial and final three-body states
involve the same three particles, the isospin coefficient simply
reduces to unity.  

%
                 \subsection{\bf Practical calculation}

For the practical calculation, we must first define the values of the
various isospins, spins and angular momenta. At first, we consider the
isospin basis case. The total isospin of the
$K^-d$ system is ${\cal I}=1/2$. Now, in a given three-body channel, the
quantum numbers $(L,S,J,I)$ of the pair and the spin $s$ of the 
spectator particle are fixed. The channel spin $\Sigma$ is then
given by: $|s-J| \leq \Sigma \leq s+J$, and the angular 
momentum $l$ of the spectator relative to the pair by: 
$|{\cal J}-\Sigma| \leq l \leq {\cal J}+\Sigma$, with $\cal J$ the
three-body total angular momentum. For a given value of $\cal J$,
the possible values of $l$ can be ordered in two sets corresponding to
opposite parities of the three-body system. The situation is summarized 
in Table \ref{tab:qn23c}.

\begin{table}[htb]
\squeezetable
\caption{Two-body ($L,S,J,I$) and three-body ($l,\Sigma,\cal J$)  
quantum numbers in the isospin basis. The two-body partial waves are 
labeled as : $^{2S+1}{\!L_J}$ for $NN$ and $YN$, $L_{2I,2J}$ for $\pi N$, 
and $L_{I,2J}$ for $\Kbar N$ and $\pi Y$.  
The column labeled as $l_a$ ($l_b$) corresponds to negative (positive)
parity states for odd values of $\cal J$, and to positive (negative)
parity states for even values of $\cal J$.
Only the values $l \ge 0$ are retained.}
\protect\label{tab:qn23c}
\begin{center}
\begin{ruledtabular}
\begin{tabular}{ccccccccc}
 channel   &     & $L$ & $S$ & $J$ & $I$ & $\Sigma$ & $l_a$ & $l_b$  \\
\hline
$\Kbar (NN)_{^3S_1-^3D_1}$ & $d$ 
                   & 0,2 & 1 & 1 &  0 & 1 & ${\cal J}+1$  & $\cal J$     \\
             &     &     &   &   &    &   & ${\cal J}-1$  &              \\
\hline
$N(\Kbar N$-$\pi\Sigma$-$\eta \Lambda)_{S_{01}}$  
                                      & $y_1$-$\alpha_1$-$\mu_1$ 
                      & 0 & 1/2 & 1/2 & 0 & 0 &              &  $\cal J$  \\
              &       &   &     &     &   & 1 & ${\cal J}+1$ &  $\cal J$  \\
              &       &   &     &     &   &   & ${\cal J}-1$ &            \\
\hline
$N(\Kbar N$-$\pi\Sigma$-$\pi\Lambda$-$\eta \Sigma)_{S_{11}}$  
                                      & $y_2$-$\alpha_2$-$\alpha_3$-$\mu_2$ 
                      & 0 & 1/2 & 1/2 & 1 & 0 &              &  $\cal J$  \\
              &       &   &     &     &   & 1 & ${\cal J}+1$ &  $\cal J$  \\
              &       &   &     &     &   &   & ${\cal J}-1$ &            \\
\hline
$\Sigma(\pi N)_{P_{33}}$  & $\Delta$ 
                  & 1 & 1/2 & 3/2 & 3/2 & 1 &    $J$       &  ${\cal J}+1$  \\
              &   &   &     &     &     &   &              &  ${\cal J}-1$  \\
              &   &   &     &     &     & 2 & ${\cal J}+2$ &  ${\cal J}+1$  \\
              &   &   &     &     &     &   &   $\cal J$   &  ${\cal J}-1$  \\
              &   &   &     &     &     &   & ${\cal J}-2$ &                \\
\hline
$\pi(\Sigma N$-$\Lambda N)_{^3S_1}$ & $\beta_1$-$\beta_2$ 
                        & 0 & 1 & 1 & 1/2 & 1 &  ${\cal J}+1$  &   $\cal J$ \\
                &       &   &   &   &     &   &  ${\cal J}-1$  &            \\
\hline
$\pi(\Sigma N)_{^3S_1}$ & $\beta_3$ 
                        & 0 & 1 & 1 & 3/2 & 1 &  ${\cal J}+1$  &   $\cal J$ \\
                &       &   &   &   &     &   &  ${\cal J}-1$  &            \\
\end{tabular}
\end{ruledtabular}
\end{center}
\end{table}

In the present paper, we consider only the $K^-d$ scattering length
which is defined as:

\begin{equation}
A_{K^-d}= - \displaystyle\lim_{p_K\rightarrow 0}
 \frac{1}{32\pi^2\sqrt s} X_{dd} ,
\end{equation}
    
\noindent
where $X_{dd}$ is the $({\cal J}=1^-, l=l'=0)$ partial amplitude for 
$K^-d$ elastic scattering, evaluated at the zero limit for 
the kaon momentum.
If we retain the contributions of the $d$+$\Kbar N$+$\Delta$+$YN$ two-body 
channels, we have a system of 12 coupled three-body channels 
(see Table \ref{tab:labels}). After angular momentum 
reduction, we obtain for ${\cal J}=1^-$ a system of 25 coupled equations, 
see Table \ref{tab:qn23c}. 
The singularities of the kernel $ZX$ are avoided by using the 
rotated contour method \cite{Heth67}, and, after discretization of the 
integrals, this system is transformed into a system of linear equations.

It is well known that the iterated form of the three-body equations does
not converge. This is illustrated with the following results obtained
(in the isopsin basis) with the "OSA+deuteron-A" model. The values 
of $A_{K^-d}$ obtained at first, second and third order of iteration 
are respectively (in fm): 
$ (0.303 + i\, 3.258) $, $ (-0.572 + i\, 3.684) $, 
and $ (-1.366 + i\, 4.131) $, which clearly do not converge.
To obtain the exact value, we solve the linear system by matrix inversion.
In the above example, we get: $ (-1.636 + i\, 2.618) $ fm.
We can also use the Pade approximants technique which leads to a convergent
solution from the successive iterated terms. In practice, we have used 
a diagonal $[5/5]$ Pade (constructed with the 11 first iterates), which
was found to be sufficient to achieve convergence. 
The dimension of the matrix to be inverted is rather large, especially
when the contributions of the "small" two-body partial waves are taken
into account. However, this is a sparse matrix  because of the 
limited number of non-zero Born terms. Thus, it is much less 
time consuming to solve the linear system with using the Pade approximants 
method.

The extension to the particle basis case is straightforward. 
As explained above, in this case the Pade approximant method must be 
even more prefered to the usual methods for solving the linear system, 
since the number of coupled channels is much larger than in the 
isospin basis for a given choice of two-body contributions, but with
a large number of Born terms equal to zero.

Note also that we have checked that, using the particle basis computer program 
with the particle masses in the multiplets replaced with the mean values used 
in the isospin basis, we have obtained again the isospin basis
results.


\section{Results}     
\protect\label{sec:results}

In this Section, we present our result for the scattering length $A_{K^-d}$.
As an introduction, we argue  why we think it necessary 
to go beyond the 
Fixed Centre Approximation (FCA) as adopted by Kamalov {\it et al.} 
\cite{Kam01}.  
Then we discuss some general aspects regarding the
choice of the basis (viz. particle vs. isospin) and the use of different 
coupled two-body $\Kbar N$ input, and we  investigate the effects due to 
the choice of different deuteron models. At the end, the effect of 
"small" two-body input on the $K^-d$ scattering length is discussed.
  
      \subsection{\bf Why need to go beyond FCA ?}
\protect\label{sec:fca}
In the Introduction, it has been stated that we need to go beyond the FCA 
and to solve the 
three-body equation exactly. We justify that claim here. 

\begin{figure}[htb]
\includegraphics[width=0.5\linewidth]{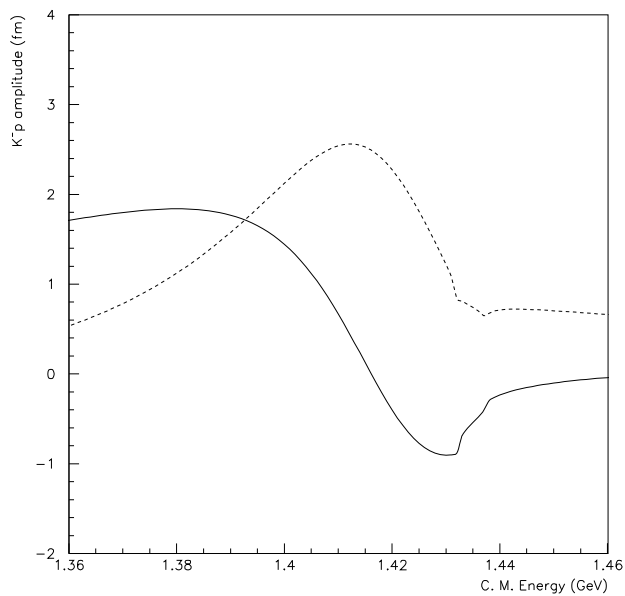}
\caption{$K^-p$ elastic scattering amplitude as a function of 
the c.m. total energy,
calculated in the particle basis with the "OSA+deuteron-A" model.
Full line: real part, dashed line: imaginary part.}
\protect\label{fig:tkp}
\end{figure}

In the FCA the deuteron is viewed as composed of a proton and a neutron  
 with a fixed separation $r$. The incoming zero energy $K^-$-meson then makes 
 multiple 
scattering off the proton and neutron  with no recoil of the target particles. 
Within this 
approximation the three-body
scattering equation may be solved algebraically: see Eq. (23) of ref.
\cite{Kam01}, to find the scattering 
length operator ${\widehat A}_{K^-d}(r)$ expressed in terms of the $\Kbar N$ 
scattering lengths: $a_p,\ a_n, \ a^\circ_n,\ a_{ex}$ as in  
Table \ref{tab:AKN}, and the separation $r$. 
The actual $K^-d$ 
scattering length is then identified as the expectation value 
$<\psi_d|\widehat A_{K^-d}|\psi_d>$ 
over $r$ with respect to the deuteron wave function $\psi_d(\vec r\,)$.  
So essentially, $A_{K^-d}$ is determined by the two-body $\Kbar N$ scattering 
lengths mentioned above.

\begin{figure}[htb]
\includegraphics[width=0.5\linewidth]{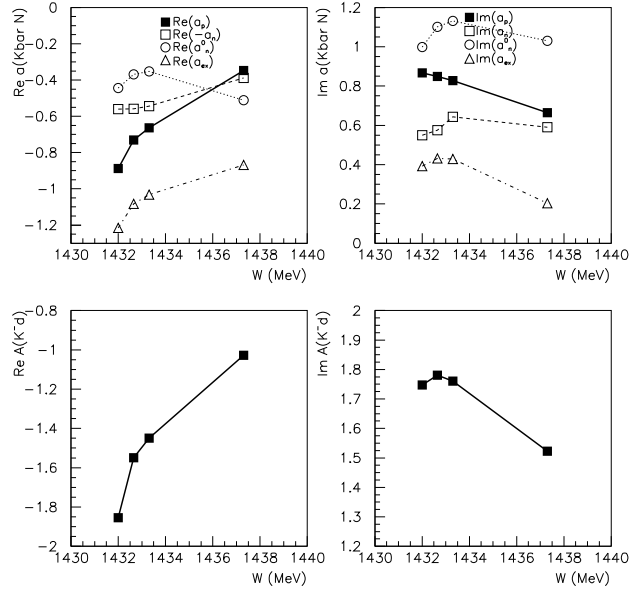} 
\caption{$\Kbar N$ and $K^- d$ scattering lengths
calculated in the particle basis for the following values of
the two-body threshold energy $W$: 
$M_{K^-}+M_p=1432$  MeV, $M_{K^-}+M_n=1432.65$  MeV,
$M_{K^-}+(M_p+M_n)/2=1433.3$  MeV, and $M_{\Kbar ^0}+M_n=1437.3$  MeV.
Model OSA is used for the $\Kbar N$ interaction. 
The $K^- d$ scattering length is calculated with the FCA approximation, 
using the "OSA+deuteron-A" model. Symbols are placed at the threshold values.
The lines are to guide the eyes.}
\protect\label{fig:knkd}
\end{figure}

As discussed in Section \ref{sec:two-body}, the  $\Lambda (1405)\ (I=0)$ 
resonance is generated as a bound state 
of $K^-p$ embedded in the $\pi Y$ continuum. Now the position of this 
resonance is fairely close to the threshold for 
$K^-p$ ($\approx 1432$ MeV) and $\Kbar^\circ n$ ($\approx 1437$ MeV), 
respectively. Thus the
elastic $K^-p$, charge exchange: $K^-p \to \Kbar^\circ n$, and hence 
the elastic 
$\Kbar^\circ n$ scattering are all affected by this resonance, and the 
corresponding amplitudes
vary rapidly near their thresholds: see Fig. \ref{fig:tkp} for the case of 
the elastic $K^-p$ 
amplitude which shows strong variations, particularly in its real part.
The question
then arises as to at which energy these amplitudes should be calculated 
to produce the 
corresponding scattering length in use for calculating $A_{K^-d}$ in FCA. 
The binding 
energy of the deuteron is $\approx 2.25$ MeV. In Fig. \ref{fig:knkd} 

\noindent
we show the values 
of the two-body scattering lengths which enter the FCA calculation and 
the resultant $A_{K^-d}$
for different values adopted for the $\Kbar N$ threshold $W$. 
We immediately notice that because of the strong variation of the 
two-body scattering lengths,
the corresponding $A_{K^-d}$ varies also rapidly with a slight shift 
in $W$: the real part in 
particular. This demonstrates that just because of the proximity to the 
$\Lambda (1405)$ 
resonance, the FCA is not very reliable. 
In a full three-body results, the quantity corresponding to $W$
is $\sqrt{\sigma}$: the energy available to any two-body amplitude
in the presence of a spectator. This is smeared out due to a loop momentum
integration. Consequently, the full three-body results do not suffer
from this excessive sensitivity. 

      \subsection{Calculations including only the $\Kbar N$ 
                  and $N N$ interactions}
\protect\label{sec:results_b}

Here we retain only the  deuteron and the $\Kbar N$ two-body
input in the three-body calculation of  the $K^-d$ scattering length. 
This means that 
in the  multiple scattering expansion of the equations only the 
following three-body states do enter: 
$K^-(pn)$, $n(K^-p)$, $n(\Kbar^\circ n)$ and $p(K^-n)$. 
Although no explicit three-body states with hyperons enter in the 
calculation in this approximation, 
there is an implicit effect from the $\pi Y$ channels through the two-body
$\Kbar N$ input. Hence it is more than the {\it single channel} approximation 
studied in Ref.\cite{Del00}.
Within the present context we first confront the results from the particle 
and isospin bases,
which makes  us keep the result with particle basis in the subsequent 
subsections. We then study the results with different $\Kbar N$ models.     
     \subsubsection{\bf Isopin and particle bases}

The isospin symmetry breaking 
effects in the $\Kbar N$ sector have been clearly demonstrated in 
Sec. \ref{sec:kbarn_inter}, by comparing the $K^- p$ observables
obtained both in the isospin and particle bases (see Table \ref{tab:bra},
and Figs. \ref{fig:sigma} and \ref{fig:spectrum}). 
In particular, upon going from the 
particle to isospin basis, we find that the magnitude of the real part of 
$a_{K^-p}$  obtained from model "OSA+deuteron-A" decreases 
by about $20 \%$ whereas the imaginary part  increases by as much as  
$30 \%$ (see Table \ref{tab:bra}). This last tendency should be due to 
the fact that
in the isospin basis (we reiterate here that in this case the isospin 
symmetry is exact) two-body channel thresholds become identical among 
different $\Kbar N$ channels within the same iso-multiplets. Consequently,  
these effects are reflected in $A_{K^-d}$, 
as one can see Table \ref{tab:AKmd_1} for different $\Kbar N$ models: 
by going from the particle to isospin basis,
the magnitude of its real part gets smaller by  about $10 \%$, while 
its imaginary part 
gets  strongly enhanced by  $\sim 70 \%$. In what follows  we will mostly 
comment on the results from the particle basis since they are more realistic. 

           \subsubsection{\bf Results using different $\Kbar N$ models}

We give in Table \ref{tab:AKmd_1} the values of $A_{K^- d}$ obtained from
models OS1, OSA and OSB of the $\Kbar N$ interactions described in
Section \ref{sec:kbarn_inter}. We briefly recall that by construction all 
the models have adopted the SU(3) symmetry constraint  on the strength 
parameters, with the possible breaking effects introduced in OSA and OSB,
and the contributions of the $\eta Y$ channels have been  taken into account, 
except for OSB.

\begin{table}[htb]
\squeezetable
\caption{Sensitivity of the $K^- d$ scattering length to the model
used for the $\Kbar N$ interaction. The deuteron is model A. The calculations 
are done in the particle and isospin bases.}
\protect\label{tab:AKmd_1}

\begin{center}
\begin{ruledtabular}
\begin{tabular}{cccc}
Model & OS1 & OSA & OSB   \\
\hline
particle basis  & $ -1.985 + i\, 1.642 $ & $ -1.802 + i\, 1.546 $ 
                & $ -1.722 + i\, 1.354 $  \\
isospin basis   & $ -1.759 + i\, 2.907 $ & $ -1.636 + i\, 2.618 $ 
	        & $ -1.709 + i\, 2.247 $  \\  
\end{tabular}
\end{ruledtabular}
\end{center}
\end{table}
The calculations have been done with  model A  for the deuteron.
The variation observed on the real and imaginary parts of $A_{K^- d}$
calculated in the particle basis are moderate: for example, the real part 
increases by about $9\%$ and the imaginary part decreases by about $6\%$, 
when replacing model OS1 by model OSA. 
The absolute values of these variations are comparable to 
those observed in the $K^- p$ scattering length (see Table \ref{tab:bra}): 
by going from OS1 to OSA
we find an increases in the real part by about $13\%$. For the
imaginary part the corresponding increase is less than $1\%$.

Concerning  the contribution from the $\eta Y$ channels
in the $\Kbar N$ interactions, we observe the same type of effects, both
on $a_{K^- p}$ and $A_{K^- d}$. Specifically, compared with the values 
given by model OSA,
both the real and imaginary parts of $a_{K^- p}$ decrease by $\sim 9\%$
when using OSB (see Table \ref{tab:bra}), while the real part of $A_{K^- d}$ 
increases by $\sim 4\%$ and the imaginary part decreases by $\sim 12\%$.
Now we recall the result in Section \ref{sec:kbarn_inter}: using model OSA 
and excluding the 
contributions of the $\eta Y$ channels,
we have seen that all $\Kbar N$ observables 
were strongly affected. In particular, the imaginary part of $a_{K^- p}$ 
increases by as much as $\sim 60 \%$, and the real part by $\sim 18 \%$ 
(see Table \ref{tab:bra}). A similar tendency is observed on $A_{K^- d}$,
as the OSA value of  $ (-1.802 + i\, 1.546) $ fm 
has been transformed to $(-1.705 + i\, 2.360)$ fm
when the $\eta Y$ channel contributions are excluded (as discussed in 
Section \ref{sec:kbarn_inter} 
this means that the
coupled channels equations were solved again with the parameters of OSA 
but excluding the
coupling to the $\eta Y$ channels).  In conjunction with the 
observation 
in Section \ref{sec:kbarn_inter} just mentioned above, we conclude here 
that
it is important to retain these $\eta Y$ channels for $A_{K^-d}$. Before 
ending this 
subsection we should mention that, as shown in our previous publication 
\cite{Baha02}, we have also tested  the $\Kbar N$ amplitude from 
Ref.\cite{Ose98} in the 
three-body equation, and the result turned out to be very close to the 
one with OS1, as expected.

	     \subsubsection{\bf Dependence on deuteron models}
		
We have tested three different deuteron models: A, B, C as described in 
Sec. \ref{sec:nn_inter}. Their $D$ state probabilities: $P_D$,  are  
$6.7\%$, $5.8\%$, and 
$0\%$, respectively.
These deuteron models are used in combination with the OS1, 
OSA and OSB parametrizations of the $\Kbar N$ interaction, and results are 
summarized in Table \ref{tab:AKmd_2}.  

\begin{table}[htb]
\squeezetable
\caption{Sensitivity of the $K^- d$ scattering length, calculated in the
particle basis, to the model used for the deuteron channel. 
Models A and B have $D$-state percentage values
of $6.7 \%$ and $5.8 \%$, respectively, and model C is pure $^3S_1$.
Models OS1, OSA and OSB of the $\Kbar N$ interactions are considered.}
\protect\label{tab:AKmd_2}

\begin{center}
\begin{ruledtabular}
\begin{tabular}{cccc}
Model & A & B & C   \\
\hline
OS1  & $ -1.985 + i\, 1.642 $ & $ -1.966 + i\, 1.515 $ 
                & $ -1.975 + i\, 1.313 $  \\
		
OSA   & $ -1.802 + i\, 1.546 $ & $ -1.788 + i\, 1.435 $ 
	        & $ -1.780 + i\, 1.243 $  \\ 
		
OSB   & $ -1.722 + i\, 1.354 $ & $ -1.703 + i\, 1.263 $ 
	        & $ -1.685 + i\, 1.128 $  \\ 
		
\end{tabular}
\end{ruledtabular}
\end{center}
\end{table}

Irrespective of which $\Kbar N$ 
model is adopted, there is a  
definite pattern: the imaginary part of the scattering length increases
in accordance with the corresponding increase in the $D$-state 
probability of the deuteron.
The increase in this quantity due to a change: $P_D=0 \to 6.7\%$, is 
as large as $20\%$.  
The real part appears to stay more or less the same in the meantime. 
We have looked at a 
few first terms in the three-body multiple scattering
series, but that does not tell us why the change is almost exclusively 
in the imaginary part. 
But whether it is in real or imaginary part, the following simple 
observation  
should suffice in 
understanding the change of this magnitude. First, we introduce the $S$ and 
$D$ component of the 
deuteron wave function in momentum space as  $\psi_S(p), \ \psi_D(p)$,
such that
\begin{equation}
P_S=\int^{\infty}_0\psi^2_S(p)p^2dp,\ \
P_D=\int^{\infty}_0\psi^2_D(p)p^2dp,\ \ P_S+P_D=1,
\end{equation}
\noindent where, for example, we may take 
$P_S \sim 0.93,\ P_D\sim 0.07$ for model A deuteron.
We may then re-write 
\begin{equation}
\psi_S(p)=\sqrt{P_S}\phi_S(p),\ \ \psi_D(p)=\sqrt{P_D}\phi_D(p),
\end{equation}
\noindent such that both $\phi_S$ and $\phi_D$ are normalized to unity. 
Note here that 
$\sqrt{P_S} \sim 0.97$ and $\sqrt{P_D} \sim 0.26$, respectively.  Then 
we may express the scattering length in the following manner:
\begin{equation}
A_{K^-d} = P_S<\phi_S|\widetilde A_{SS}|\phi_S>
          +2\sqrt{P_D P_S}<\phi_D|\widetilde A_{DS}|\phi_S>
          +P_D<\phi_D|\widetilde A_{DD}|\phi_D>,
\end{equation}
\noindent 
where $\widetilde A$ are operators in the space of the deuteron
wave function, and
the closed brackets mean the integration over the initial and final
loop (or off-shell) momenta. In the above expression, it is easy to see 
that due to the second $S$-$D$ {\it interference}  term the 
result with and without the deuteron $D$ state could differ up to  
a few 10's of a percent, even though $P_D$ is just about $\sim 6\%$. 
Of course it is likely that the matrix element 
$<\phi_D|\widetilde A_{DS}|\phi_S>$ 
may well be smaller than the one for the first term: 
$<\phi_S|\widetilde A_{SS}|\phi_S>$,
due to a slight angular momentum mismatch between the $S$ and $D$ states. 
However,
when we study the behavior of $\psi_S(p)$ and $\psi_D(p)$, it is easy 
to observe the 
following trend: the former is very large at $p=0$ and decreases 
rather rapidly down to
$p \equiv 0.6 \ \hbox{fm}^{-1}$. The latter is zero at $p=0$ to start with,
but it's magnitude increases
rapidly up to about $p=0.6\ \hbox{fm}^{-1}$, then decreases moderately down to
about $p=2.5\ \hbox{fm}^{-1}$.  
The result is that the wave function components become of the same 
order of magnitude 
from about
$p_{equal}=0.75\ \hbox{fm}^{-1}$. On the other hand, we find that $p_{equal}$ 
is in the 
sub-threshold region for the $K^-p$ amplitude  dominated by the 
$\Lambda (1405)$
resonance, see Fig. \ref{fig:tkp}.  This is imbedded in 
$\widetilde A_{DS}$.
  For this reason 
it may be fairly likely that 
$<\phi_D|\widetilde A_{DS}|\phi_S>$ and
$<\phi_S|\widetilde A_{SS}|\phi_S>$  are not very different in magnitude, 
hence a $\sim 20\%$ 
increase  in the imaginary part of $A_{K^-d}$ due to the deuteron
 $D$ state is possible. 
So it is important to retain that component.

      \subsection{Effects of the "small" two-body input}
\protect\label{sec:results_c}

What we term here as small two-body input are (i) the $\pi Y$ channels 
resulting from the initial  $\Kbar N$ interaction,  
(ii) $\pi N$ appearing with a spectator hyperon, and (iii) the $YN$ 
interactions with a spectator nucleon in the three-body equations. 
Here we exclude the channels involving the $\eta$ meson, 
recall Tables \ref{tab:labels}. In this manner, in principle, our equations 
do satisfy three-body unitarity exactly.  
It was shown in Refs. \cite{Tok81,Tor86} that these interactions were
very important in the threshold break-up reactions: $K^-d \to \pi NY$ 
as they control the final state interactions.  Here we want to study 
these effects in the threshold elastic case. For some convenience, 
calculations in the isospin basis come back in our discussion. 
   
We take models OS1 and OSA for the $\Kbar N$ interactions, and model A
for the deuteron.  Then, we add successively
the $\pi N$-$P_{33}$ and $Y N$ interactions described in
Sec. \ref{sec:other_input}. The results are given 
in Table \ref{tab:AKmd_3}: the third and fourth columns give the values 
obtained when only the $\pi N$-$P_{33}$ or $Y N$
input are added to the $\Kbar N$+$d$ input, and in the last column
both contributions are taken into account. 

\begin{table}[htb]
\squeezetable
\caption{Contributions of the small two-body input to the $K^- d$ 
scattering length (values in fm). Models OS1 and OSA are used for
the $\Kbar N$ interaction, and model A for the deuteron. 
The calculations are done in the particle basis.}
\protect\label{tab:AKmd_3}

\begin{center}
\begin{ruledtabular}
\begin{tabular}{ccccc}
Model & $d+\Kbar N$ & + $\Delta$ & + $Y N$ 
      & + $\Delta$ + $Y N$  \\
\hline
OS1  & $ -1.985 + i\, 1.642 $ & $ -1.985 + i\, 1.663 $ 
	        & $ -1.975 + i\, 1.611 $ & $ -1.974 + i\, 1.634 $ \\  
OSA  & $ -1.802 + i\, 1.546 $ & $ -1.793 + i\, 1.562 $ 
                & $ -1.805 + i\, 1.511 $ & $ -1.796 + i\, 1.529 $ \\
\end{tabular}
\end{ruledtabular}
\end{center}
\end{table}

The effect
of the additionnal two-body input is negligible. This can be easily
understood by  explicitely writing the three-body equations. For example,
let us consider Eq. (\ref{eq:3body}) written in the isospin basis, with the 
following simplified two-body input: 
$d$ + $\Kbar N$ (limited to $I=0$, without the $\eta \Lambda$ channel 
contribution) + $\pi N$-$P_{33}$. Using the channel labels
as defined in Table \ref{tab:labels} (with $y_1$, $\alpha_1$ simplified
into $y$, $\alpha$, respectively), the explicit form of the coupled
three-body equations is:

\begin{eqnarray}
\label{eq:3-bod_simp}
\left\{
\begin{array}{llll}
 X_{dd} & = & Z_{dy} R_{yy} X_{yd} + Z_{dy} R_{y\alpha} X_{\alpha d}\\
X_{yd} & = & Z_{yd} + Z_{yd} R_d X_{dd} + Z_{yy} R_{yy} X_{yd}
                                   + Z_{yy} R_{y\alpha} X_{\alpha d}\\
X_{\alpha d} & = & Z_{\alpha \Delta} R_{\Delta} X_{\Delta d}\\
X_{\Delta d} & = & Z_{\Delta\alpha} R_{\alpha y} X_{yd}
                 + Z_{\Delta\alpha} R_{\alpha\alpha} X_{\alpha d}
\end{array}
\right.
\end{eqnarray}

Using the last equation to express $X_{\alpha d}$ in the first 
three equations, we see that the $\pi N$-$P_{33}$ channel contributes
in terms of second or higher order, thus its effect
on the low energy $K^- d$ observables like $A_{K^- d}$ should be small, 
although the resulting $\pi N$-$P_{33}$ state in the absence of a 
spectator $Y$ is nearly on the $\Delta$ 
resonance peak in its two-body center of mass energy. This is consistent 
with a semi-quantitative
estimate of the effect by Kamalov {\it et al.} \cite{Kam01}.
Similar arguments hold for the contribution of the $Y N$ interactions,
and also when the particle basis is used. 

Now, we may need to discuss the problem associated with  the fact that 
the signs of the off-diagonal parameters of the $\Kbar N$ and $Y N$ 
interactions are undetermined. 
Let us consider for example the OSA model of the $\Kbar N$ interaction. 
As explained in Section \ref{sec:kbarn_inter}, 
the signs of the off-diagonal strengths are those of the SU(3) coefficients
given in Tables II and III of Ref. \cite{Ose98}. Now, if these signs are 
changed, the signs of the corresponding off-diagonal two-body propagators
are also changed, but the $\Kbar N$ observables are not (except 
the signs of the corresponding scattering lengths). For example,
changing the signs of $\lambda_{\Kbar N-\pi \Sigma}$ both for
$I=0$ and $I=1$, and/or the sign of $\lambda_{\Kbar N-\pi \Lambda}$
(which contributes only for $I=1$), does not affect the observables.
Similar conclusions hold for the $Y N$ interactions when the sign
of $\lambda_{\Sigma N-\Lambda N}$ is reversed.
Now, the situation is not so simple in the three-body sector. To examine 
what happens, we consider the following cases in the isospin basis 
(we choose the isospin basis for the sake of simplicity in handling labels,
but the conclusions are the same in the particle basis):

{\it i)} Only the $d$+$\Kbar N$ interactions (without the $\eta Y$ channels 
contributions) are adopted. The corresponding three-body equations 
have the form (\ref{eq:3-bod_simp}), with 
$Z_{\alpha \Delta}=Z_{\Delta\alpha}=0$. Thus
Eqs. (\ref{eq:3-bod_simp}) reduce to the first two equations with 
$X_{\alpha d}=0$. As these equations do not involve the off-diagonal
$\Kbar N$-$\pi Y$ propagators, the reaction amplitudes do not depend
on the sign of the off-diagonal $\Kbar N$-$\pi Y$ coupling constants.

{\it ii)} Next, we add  the $\pi N$-$P_{33}$ two-body 
channel to the one just mentione above. Going back to 
Eqs. (\ref{eq:3-bod_simp}), 
it is clear that
we obtain the same sytem of equations if we change the signs of both
the off-diagonal $\Kbar N$-$\pi Y$ propagators and the  $X_{\alpha d}$
and $X_{\Delta d}$ amplitudes. Therefore, the $K^- d$ scattering length
will not be affected when changing the signs of the  off-diagonal 
$\Kbar N$-$\pi Y$ coupling constants.

{\it iii)} Now, we add $YN$ interactions to the model with 
the $d$+$\Kbar N$ interactions, and  change the sign of the 
$\Sigma N$-$\Lambda N$ coupling constant. Then, contrary to the previous case,
the $K^- d$ scattering length gets changed. This can be understood 
by noting that, due to isospin conservation, the $\Lambda$ exchange 
in the three-body sector is possible only between
the $\pi(\Lambda N)_{I=1/2}$ and $N(\pi \Lambda)_{I=1}$ states
(Born term $Z_{\beta_2 \alpha_3}$, see Table \ref{tab:born-terms}).
This "dissymmetry" (comparing with the situation for the $\Sigma$ exchange)
implies that we cannot change simultaneously the signs of the
$\Sigma N$-$\Lambda N$ propagators and of some of the three-body amplitudes
without changing the original system of equations. 
However, as explained before, the change in $A_{K^- d}$ does not exceed
a few percent (see Table \ref{tab:AKmd_3}), therefore this problem will
not be regarded as an important issue worthy of extensive discussion in the 
present paper. 
It should be appropriate to stress in this regard that we do not anticipate 
any significant lack 
of precision because  we have adopted a separable rank one form 
for the hyperon-nucleon 
$S$-wave interactions (which are the part of the 
``small'' input): we have clearly witnessed that these channels have been 
found to give only a  small effect 
in the calculation of $A_{K^-d}$. In particular, if  we accept that 
the $YN$ interactions are dictated
by $SU(3)$ symmetry, just like our $\Kbar N$ two-body input, and thus 
adjust all 
the signs of strengths in our separable potentials, for example,  
to the corresponding 
$S$-wave projected Nijmegen meson exchange potentials \cite{Rij99}, 
then the sign 
ambiguity will also be gone out of our discussion. In fact the related 
problem was
already studied by Dalitz et al., Ref. \cite{Tor86}, who found important
variations in the $\Lambda p$ mass spectrum in the threshold break up
reaction: $K^-d \to \pi NY$.  This will necessitate us to re-examine the 
break up channels within our present approach.  At present, we assess 
that the effect of the "small" two-body input is actually not important 
for the calculation of $A_{K^-d}$.


\section{Discussion}
\protect\label{sec:disc}

In this Section we  present the best estimate for the 
theoretical value for $A_{K^-d}$ in our three-body approach. Then we discuss
what kind of uncertainty should be associated with this value which derives
from the effects we have left out in the present calculation.

From the result presented in the last Section, we make the following choice 
for our preferred value:  the one from the combination of the two-body 
input "OSA+deuteron-A", in the particle basis, 
see Tables \ref{tab:AKmd_1} and \ref{tab:AKmd_2} 
for the all set of the results. We have not adopted the values from  
the calculation explicitly incorporating the hyperon channels which 
are associated with what we called 
{\it small two-body input} in Sec. \ref{sec:two-body}: as should be clear 
from Table \ref{tab:AKmd_3}, 
the effects are found to be quite small ($< 3\%$). Besides, effects 
due to the {\it sign ambiguity} in the off-diagonal $YN$ amplitudes 
do not exceed $1\%$. So we take a conservative estimate of 
the $K^-d$ scattering length to be  

\begin{equation}
A_{K^-d}=(-1.80 + i\ 1.55) \ \mbox {fm},
\end{equation}

\noindent to which we may assign a possible uncertainty of a few percent.

We then need to assess the effects which may not have been taken into account 
in an ordinary three-body equations approach like
the present one. The first such processes are
possible four-body intermediate states:
those with two mesons and two baryons. Diagrammatically, they may be divided 
into partially and totally connected ones. Of partially connected diagrams, 
those associated with
baryon self energies should be dropped from consideration since we assume 
to have been dealing with 
the initial $\Kbar d$ channel. Then the remaining partially connected 
diagrams are (i) the ones in which there is a spectator meson and 
two baryons exchanging a meson, and
(ii) the ones with a spectator baryon, and a meson and a baryon exchanging 
a meson.
For the first ones they 
have already been included effectively in the input baryon-baryon 
($NN$ or $YN$) interactions. Likewise,
the second ones are effectively included in the coupled $\Kbar N$ 
and $\pi N$ input since they have 
been fitted to data.  So we have only to worry about the 
completely connected diagrams.   
Quite fortunately, except practically for a couple of diagrams
\footnote{Those may be classified as meson exchange current terms 
that are found as diagrams (b) and (c) in the work of Weinberg 
\cite{Wei92} for 
the case of the pion-deuteron scattering length. This combination 
has been calculated
earlier by Robilotta and Wilkin \cite{Rob78} and found to contribute 
less than 
$3\%$ to the total scattering length. Weinberg's calculation did confirm 
this earlier result. 
We may conclude that contribution from the corresponding diagrams for
 $A_{K^-d}$ be far 
smaller.  This is based upon the forthcoming discussion in the main text, 
on the comparative study of 
$A_{K^-d}$ and $A_{\pi^-d}$  related to the meson absorption effect. 
Note that as explained in
Ref. \cite{Wei92}, a combination of diagrams (d) and (e) should give 
no contribution for the
scattering length.}
, they  are reduced to two 
baryon ($YN$) interactions: crossed two meson exchanges, and
one meson exchanges with meson-baryon-baryon vertex corrections due 
to virtual meson creation and absorption across the vertex.  
So those completely connected diagrams are just the 
pure two-body intermediate $YN$
channels resulting from  absorbing  $\Kbar$ (or $\pi$). 

We thus should consider only  the effects of the meson ($K^-$) 
absorption in the 
$K^-d $ elastic scattering at threshold. 

In Ref. \cite{Jid02}  the $p$-wave effect in the low energy $\Kbar N$ 
interaction has been studied within the context of an effective chiral 
Lagrangian. In that work, the effect derives from the $s$-channel 
pole contributions (the absorption/re-emission of $\Kbar$ by a 
nucleon: $\Kbar N \to Y \to \Kbar N$). 
So this could be used as a measure for the meson absorption effect 
under consideration.    
The authors have stated that the $p$-wave effect is quite small: 
total cross sections change very little, whereas the differential 
cross sections have improved to follow
the trend seen in  experimental data. From this publication, 
what we could  possibly 
exploit  as the indicators of  the $K^-$ absorption effect 
semi-quantitatively may be
the bare mass of $\Lambda$, viz. $\widetilde M_{\Lambda}$ or the ratio 
$R_c$, recall Section \ref{sec:two-body}
for discussion on the available experimental data for the coupled 
$K^- p$ channels. In this regard, 
as found in Table \ref{tab:bra}, the other two ratios: $\gamma$ and $R_n$ 
are too sensitive to be used for our
objective. We have observed that the change in $R_c$ by the $K^-$ absorption 
effect (viz. the inclusion of the $p$-wave in the language of 
Ref.\cite{Jid02}) is at most 
$\sim 2\%$, whereas the shift from the bare mass to the physical one due 
to the same effect for the 
$\Lambda$ is about $3\%$ (the corresponding values following Eq.(26) in that 
publication cannot be used
since the basis model for the $s$-wave $\Kbar N$ interaction has been 
modified by readjusting the subtraction constants in Eq. (26)). Since 
the kinetic energy available to the $YN$ system after the
 $K^-$ absorption in the $K^-d$ system at threshold corresponds to 
$ p^{lab}_Y \ge 1$ GeV/c, the effect of the 
$YN$ interaction is expected to be small (a reasonable guess may be reached 
from Fig. 3 in the 
present article and in Ref.\cite{Rij99}). In such a situation what we have 
just estimated above may well be interpreted as the effect of 
the $K^-$ absorption.  To be on the conservative side we set this to be 
a possible correction of a few percent. 
  
Not directly applicable but rather useful information regarding the effect 
of meson absorption comes from the $\pi^-d$ scattering length: $A_{\pi^-d}$. 
With the exception of some small effects from 
$\pi^-d \to \pi^\circ nn$ and $\pi^-d \to \gamma nn$, the scattering length 
in this process is purely real if no strong pion absorption effect 
is in effect, viz. no imaginary part in the absence of pion absorption. 
With this in mind, 
ealier model calculations indicated that
the effect creates contributions both to real and imaginary parts of the 
scattering length. An earlier three-body model calculations 
\cite{Afn74} obtained 
${\cal R}e(A_{\pi^-d})\equiv -0.035$, and ${\cal I}m(A_{\pi^-d})$ 
to be between $0.0062$ and $0.0075$, both in units of the 
inverse pion mass: $m^{-1}_{\pi}$. What should be emphasized here is 
that the pion absorption 
contributes to the real part (commonly termed as the dispersive effect) with 
just about the same magnitude as the imaginary part, but with 
the negative sign. This characteristic feature was confirmed by a 
later calculation in multiple scattering in Ref.\cite{Miz77}, 
and the pion absorption contribution was evaluated to be
\begin{equation}
\Delta A^{abs}_{\pi^-d} \equiv (-0.008 +i\ 0.011)\  m^{-1}_{\pi}. 
\end{equation}

\noindent A more complete three-body calculation explicitly including 
the pion absorption 
in a fully consistent manner \cite{Fay80} obtained
\begin{equation}
 A_{\pi^-d} = (-0.047 +i\ 0.0047)\  m^{-1}_{\pi}. 
\end{equation}
\noindent  The pion absorption may be seen as contributing roughly 
  $10\%$ to the real part.

A couple of recent papers have reported an extraction of $A_{\pi^-d}$ 
from the pionic
deuterium atomic transitions, Refs. \cite{Sch01,Cha95,Hau98}, 
 using the Deser-Trueman formula \cite{Deser} 
 to find (Coulomb interaction included): 
\begin{equation}
A^c_{\pi^-d}=(-0.0259 \pm 0.0011) + i\ (0.0054 \pm 0.0011),
\end{equation}
\noindent from Ref.\cite{Sch01,Cha95}, and
\begin{equation}
A^c_{\pi^-d}= (-0.0261 \pm 0.0005) + i\ (0.0063 \pm 0.0007),
\end{equation}
\noindent from Ref. \cite{Hau98}, both in units of the inverse pion mass 
as before. Both of these data are 
consistent with  each other, and particularly the imaginary parts are also 
consistent with the model calculations in Refs. \cite{Afn74,Fay80}.  
However, the real parts are  about 
half the magnitude of the model result in Ref. \cite{Fay80}. If we assume that
the experimental result be correct and that
the result of the model calculation mentioned above be also correct 
regarding the
sign and size of the absorption contribution to the real part, 
then the real part of the scattering length without the pion  absorption 
effect should be about $-0.020 \ m^{-1}_{\pi}$. Then 
it appears that the data indicate the  pion absorption 
effect to be very close to $30\%$.  And supposing that we translate this to 
our present $K^-d$ scattering length, the pure three-body result could not be 
acceptable.  But there is a possible way out of this impasse:
in Ref.\cite{Sch01,Cha95}, by combining the data from  the pionic hydrogen and 
pionic deuterium, the value of the scattering length $\pi^- n$, or more 
precisely the isoscalar combination:
\begin{equation}
2b^c_0=a^c_{\pi^- p}+a^c_{\pi^- n},
\end{equation}   
\noindent turned out to be consistent with zero. This {\it near} vanishing 
of $b^c_0$  should be expected 
from  current algebra calculations \cite{Wei66}, and particularly in the 
soft pion 
(viz. zero pion mass) limit (however, we should be 
reminded in this respect that the extraction of this latter quantity is 
still rather 
{\it model dependent},
see Ref. \cite{Sch01,Cha95} for details, as well as the consequence from 
the $\pi N$ 
partial wave analyses, see for example Ref. \cite{Arn95}). So if one 
accept this result, the lowest order pion-deuteron scattering length
in a static calculation vanishes, and this is the basic origin of the
smallness of ${\cal R}e (A_{\pi^- d})$. Consequently, one might well come up 
with that
large pion absorption effect.  

Now,  this is far from true 
in the case of the $K^-d$ scattering length
where (i) because the kaon cannot be regarded as {\it soft}, and 
(ii) because of the predominantly 
exo-energetic nature of the associated coupled channels,  even 
the lowest order scattering length
is neither vanishingly small nor purely real to start with, 
even without $K^-$ absorption.
Hence we may safely abide by the estimate of 
the kaon absorption effect as 
discussed earlier, and so we set the effect to be less than $10\%$.  
Of course a more quantitative study will have to be done.  

Our present calculation has not included any electromagnetic interactions,
 of which the Coulomb interaction plays the dominant role in the actual 
 hadronic quantities measured.  So 
we now come to discuss the effect of the Coulomb interaction as our last 
subject for this Section.
 Here just like what we have done above, we will borrow a good part of our 
argument below from the $\pi^-p$ and $\pi^-d$ scattering length problems. 
In fact, as long as the 
aspect related to the electromagnetic interactions is concerned, 
replacing  $\pi^-$ by  $K^-$ should not alter it in an essential manner.  
So the first important point to be stressed is the following: 
all the experimental determination  of the $\pi^-p$ elastic  
(and $\pi^-p  \to \pi^\circ n$
charge exchange) scattering lengths extracted from the pionic hydrogen 
atom level shift 
and width have taken care of various electromagnetic corrections 
(including the finite
electromagnetic size of the pion, vacuum polarisation effect, etc.) to the 
Deser-Trueman formula only \cite{Sch01,Cha95,Hau98}.  
As found, for example, in Ref.\cite{Sig96}, this correction is up to 
about $2\%$. But 
the effect of the point Coulomb interaction, which is the very basis 
for the use of the
Deser-Trueman formula,  has not been taken out. Thus to be 
more precise, the extracted quantities
should carry an index ``c'' to indicate that the Coulomb effect is 
still there. This we 
have done explicitly in what we have written above. In the case of 
the pion-deuteron 
scattering length, even this type of electromagnetic corrections
to the Deser-Trueman formula has not been attempted. To a large 
extent the reason should be that 
the calculation is far more complicated than for the $\pi^-p$ case. 
But one may well suspect that 
a straightforward application of the Deser-Trueman formula  to this 
already explicitly extended
system would obviously introduce an error far larger than these 
sophisticated correction to that very  formula. 

There have been model dependent but rather detailed calculations 
relating $a^c_{\pi^-p}$ and 
$a_{\pi^-p}$, the latter being due to the purely strong interaction 
\cite{Del01,Del03}.
It was stated that the difference between the two quantities is just 
a fraction of a percent.
As the same method cannot be applied,  a very simply  estimate was 
carried out to 
relate the corresponding quantities for the $\pi^-d$ scattering length 
\cite{Del01}:
\begin{equation}
A^c_{\pi^- d}/A_{\pi^- d} =\int^{\infty}_0 u^2_{Deu}(2r) \phi^2_0(0,r) dr
/\int^{\infty}_0 u^2_{Deu}(2r) r^2 dr,
\end{equation}
\noindent where $u_{Deu}(r)$ is the $S$-wave radial deuteron wave 
function and 
$\phi_0(0,r)$ is the $S$-wave zero momentum Coulomb wave function for 
a unit charge. 
Clearly, this is just to semi-quantitatively introduce the distortion 
of the incoming 
charged pion due to Coulomb interaction.  The result is about $4\%$ 
increase in the 
magnitude for $A_{\pi^-d}$ as compared with its Coulomb included 
counterpart.  However, 
as we see in the experimentally extracted $A^c_{\pi^-d}$ reported above 
\cite{Cha95,Hau98}, the error bars are just about the size estimated here. 
So here again, the Coulomb correction seems to appear quite small. 
Likewise,  the
same line of reasoning might well apply to the case of the $K^-d$ 
scattering length.
 When translated into the model prediction, a possible  allowance 
 should be
taken into consideration between the purely strong and Coulomb included 
scattering lengths, although by its very nature the estimate 
should be regarded qualitative.
To this end it should be useful to refer to the work of Barrett and Deloff 
\cite{Bar99}. They introdued a set of rather simple $K^-d$ optical 
potentials and 
calculated  the strong interaction  shift and width of the $1S$  
atomic level for the kaonic 
deuterium.  Also the optical potentials were used to calculate the 
purely  strong ($A_{K^-d}$), as well as the Coulomb included ($A^c_{K^-d}$)
$K^-d$ scattering lengths. The observation they made was that the 
Deser-Trueman
formula might be inaccurate, and that a blind application of that 
formula and the identification of the 
extracted quantity as the $K^-d$ scattering length due only to 
strong interaction might 
introduce an error as large as $20\%$. A word of caution should be due 
regarding this work:
the optical potentials constructed there were quite simple, and strong
 non-locality
expected from the dominance of the $\Lambda(1405)$ was absent. So a 
more realistic 
optical potential should be constructed in order to give more reliable
statements on the issues. Otherwise, if one wants to simply obtain the 
Coulomb included
$K^-d$ scattering length, it is possible to use the pure-strong $K^-d$ 
amplitude and apply Coulomb corrections as found, for example, in  
Ref.\cite{Fro85}.


\section{Summary and Conclusions} 
The scope of the present work being to develop a reliable formalism to 
calculate the 
$K^-d$ scattering length, our starting point was a thorough study of various 
appropriate two-body processes. For that purpose, we first focused on 
the highly 
inelastic $K^-p$ initiated reactions in the kinematics region 
$p^{lab}_K\le$ 250 MeV/c. 
The elastic, as well as the relevant seven inelastic coupled-channels were 
investigated via an effective non-linear chiral meson-baryon Lagrangian. 
Within a broken SU(3)-symmetry scheme, the adjustable parameters of the 
formalism were determined by a fitting procedure on the threshold branching 
ratios and total cross-section data, leading to reduced $\chi^2$'s 
close to 1.2.

To make clear the sensitivity of the observables to the phenomenological 
ingredients, 
three models were constructed. They were then exploited to predict other 
measured 
quantities, namely, 

\begin{itemize}
\item{
$K^-p$ scattering length, for which our best value is 
$$a_{K^-p} ~=~ ( -0.90  + i\ 0.87 )\ \hbox{fm},$$
\noindent 
in agreement, within the experimental uncertainties, with the 
recent KEK data,
$$a^c_{K^-p} ~=~      (-0.78 \pm 0.15  \pm 0.03) 
           + i   (0.49 \pm 0.25  \pm 0.12)\ \hbox{fm}.$$
}
\item{The $\Sigma \pi$ mass spectrum, measured at CERN some 20 years ago, 
was reproduced 
in an acceptable manner. This quantity was found quite sensitive to 
the $\eta Y$ 
intermediate state within the used coupled-channel approach.
}
\end{itemize}

In view of the $K^-d$ system investigations, besides the $K^-p$ interactions, 
one of course 
needs another elementary amplitude: $K^-n$, for which no data is available. 
We hence 
performed predictions for the elastic and the four inelastic coupled-channels. 
Moreover, predictions for various scattering lengths were made, i.e.,
$a_n(K^-n \to K^-n)$, $a_n^\circ(\Kbar^\circ n \to \Kbar^\circ n)$, 
and $a_{ex}(K^-p \to \Kbar^\circ n)$.
Finally, the implementation of the two-body sector was completed by 
studying other relevant 
channels: pion-nucleon, nucleon-nucleon (the deuteron), and 
nucleon-hyperon interactions. 

Then, we moved to the central topic of the present work and developed a 
relativistic 
version of the three-body Faddeev equations, to which we embodied the above 
elementary 
operators. As expected, this formalism allows 
us to go far beyond previous investigations, such as single- and 
multi-channel approaches 
and fixed center approximation.

As in the case of $\Kbar N$ interactions, we performed our studies in 
both isospin 
and particle bases.

Investigating the $K^-d$ system brings in phenomena with small contributions. 
A quantitative determination of their importance goes beyond the scope 
of the present work. 
However we tried to evaluate them qualitatively, i.e., possible contributions 
from four-body 
intermediate states, $K^-$ absorption, and Coulomb correction, this latter 
being also 
present in the elementary two-body channels.

Finally, our best value for the $K^-d$ scattering length is,
$$A_{K^-d} =     ( -1.80  + i\ 1.55 )\ \hbox{fm}.$$

Given the quality of the phenomenological input and approximations 
introduced, our 
estimations lead us to attribute to the above values an uncertainty 
of about 10\%.

The awaited for data will soon make clear how realistic our predictions are. 
These experimental results will come from $DA \Phi NE$ on the 
$\Kbar N$ and $K^-d$ scattering lengths, 
as well as from COSY, ELSA, JLab, and SPring-8 on the lowest mass 
$\Lambda$-resonances, 
including the $\Sigma \pi$ mass spectrum.
From theoretical side, several topics deserve to be studied, such as 
Coulomb effects and $K^-d$ break-up channels.

\begin{acknowledgments}

A. B. would like to thank DSM/DAPNIA, CEA/Saclay, and IPN-Lyon for 
their kind hospitalities. 
T. M. would like to acknowledge  a very pleasant hospitality extended to 
him at IPN-Lyon 
where the final phase of the present work was performed. We are greatful 
to Angels Ramos 
for informative correspondences on some of her works pertinent to the 
present one. 
Thanks go also to Andrej Deloff  for discussions on the subject of Coulomb 
interactions in low energy hadronic systems.     
\end{acknowledgments}

\begin{appendix}

\section{}
\protect\label{app-a}
\centerline{\bf Particle and isospin basis}
We consider two particles with isospin $I_1$, $I_2$, projections
$I_{1z}$, $I_{2z}$, and total isospin $I$, $I_z$.
The transformation from isospin to particle basis is written as:
\begin{equation}
\ket{I_1 I_{1z} I_2 I_{2z}} = 
       \sum_{I=|I_1-I_2|}^{I_1+I_2} 
       \bra{I_1 I_{1z} I_2 I_{2z}} \,I I_{1z}+I_{2z} \,\rangle \>
              \ket{I I_{1z}+I_{2z}}.
\label{eq:isophys}
\end{equation}

Next, we specify the values of the isospins and their projections of
the particles, following the phase convention given in Table \ref{tab:masses}
for the isospin states. Calculating the appropriate Clebsch-Gordan 
coefficients, we easily obtain the linear relations between the states in
the two bases. 
         \subsection{$\Kbar N$ states}
For the $K^- p$ interactions, the eight physical states defined in 
Eqs. (\ref{eq:ch1-8}) are expressed as linear combinations
of the $I=0$, $1$ and $2$ states, according to the following relations:
\begin{eqnarray*}
\left\{
\begin{array}{llllllll}
\ket{K^- p} &=& \frac{1}{\sqrt{2}}\,
              \left[\ket{0\,0}_{\Kbar N} - \ket{1\,0}_{\Kbar N} \right] , \\
%
%
\ket{\Kbar^0 n} &=& \frac{1}{\sqrt{2}}\,
                  \left[\ket{0\,0}_{\Kbar N} + \ket{1\,0}_{\Kbar N}\right] , \\
%
%
\ket{\pi^- \Sigma^+} &=& 
       - \frac{1}{\sqrt{3}}\,\ket{0\,0}_{\pi \Sigma} +
         \frac{1}{\sqrt{2}}\,\ket{1\,0}_{\pi \Sigma} - 
	 \frac{1}{\sqrt{6}}\,\ket{2\,0}_{\pi \Sigma} , \\
%
%
\ket{\pi^+ \Sigma^-} &=& 
       - \frac{1}{\sqrt{3}}\,\ket{0\,0}_{\pi \Sigma} -
         \frac{1}{\sqrt{2}}\,\ket{1\,0}_{\pi \Sigma} - 
	 \frac{1}{\sqrt{6}}\,\ket{2\,0}_{\pi \Sigma} , \\
%
%
\ket{\pi^0 \Sigma^0} &=& 
       - \frac{1}{\sqrt{3}}\,\ket{0\,0}_{\pi \Sigma} +
         \sqrt{\frac{2}{3}}\,\ket{2\,0}_{\pi \Sigma} , \\
%
%
\ket{\pi^0 \Lambda} &=& \ket{1\,0}_{\pi \Lambda} , \\
%
%
\ket{\eta \Sigma^0} &=& \ket{1\,0}_{\eta \Sigma} , \\
%
%
\ket{\eta \Lambda}  &=& \ket{0\,0}_{\eta \Lambda} .
\end{array}
\right.
\end{eqnarray*}

From these expressions, we can deduce the relations between the transition
potentials in the two bases.
Choosing a separable form as Eq. (\ref{eq:separable}), with isospin-independent 
form factors, we obtain the relations between the strength 
parameters in the two bases. Neglecting the contributions of the $I=2$ states,
we have for example:

\begin{equation}
\label{eq:a2}
\lambda_{K^- p - \Kbar^0 n} = 
    \frac{1}{2} [ \lambda_{\Kbar N-\Kbar N}^0 -
                  \lambda_{\Kbar N-\Kbar N}^1 ]  \quad , \quad 
\lambda_{\Kbar^0 n - \pi^+\Sigma^-} =
   -\frac{1}{\sqrt 6}\lambda_{\Kbar N-\pi \Sigma}^0 + 
    \frac{1}{2}\lambda_{\Kbar N-\pi \Sigma}^1         \quad , 
\end{equation}

\noindent
and so on for the other parameters, with the symmetry property: 
$\lambda_{ij} = \lambda_{ji}$.

In the case of the $K^- n$ and related states, only the $I=1$ and $2$ states 
contribute, and the five physical states Eq. (\ref{eq:ch1-5}) are expressed 
as follows:

\begin{eqnarray*}
\left\{
\begin{array}{lllll}
\ket{K^- n} &=& - \ket{1\,-1}_{\Kbar N} , \\
%
%
\ket{\pi^- \Sigma^0} &=& 
       - \frac{1}{\sqrt{2}}\,\ket{1\,-1}_{\pi \Sigma} +
         \frac{1}{\sqrt{2}}\,\ket{2\,-1}_{\pi \Sigma} , \\
%
%
\ket{\pi^0 \Sigma^-} &=& 
         \frac{1}{\sqrt{2}}\,\ket{1\,-1}_{\pi \Sigma} +
         \frac{1}{\sqrt{2}}\,\ket{2\,-1}_{\pi \Sigma} , \\	 
%
%
\ket{\pi^- \Lambda} &=& \ket{1\,-1}_{\pi \Lambda} , \\
%
%
\ket{\eta \Sigma^-} &=& \ket{1\,-1}_{\eta \Sigma} .
\end{array}
\right.
\end{eqnarray*}

The relations between the strength parameters in the two bases are obtained 
along the same lines as in the $K^- p$ case. Neglecting as before the
$I=2$ states contributions, we obtain for example:

\begin{equation}
\lambda_{K^- n - K^- n} = \lambda_{\Kbar N-\Kbar N}^1 \quad , \quad
\lambda_{K^- n - \pi^- \Sigma^0} = 
           \frac{1}{\sqrt 2}\lambda_{\Kbar N-\pi \Sigma}^1 .
\end{equation}
 
\noindent
etc.

               \subsection{$\pi N$ states}

The different $\pi N$ states in the particle basis are expressed as
linear combinations of the $I=\frac{1}{2}, \frac{3}{2}$ isospin
states. Considering the possible charge states of the pion and nucleon,
we have the following two sets of coupled states:

\begin{eqnarray*}
\left\{
\begin{array}{ll}
\ket{\pi^0 p} = 
         \sqrt{\frac{2}{3}} \,\ket{\frac{3}{2}\, \frac{1}{2}}_{\pi N} -
	 \frac{1}{\sqrt 3}  \,\ket{\frac{1}{2}\, \frac{1}{2}}_{\pi N} , \\
%
%
\ket{\pi^+ n} = 
       - \frac{1}{\sqrt 3}  \,\ket{\frac{3}{2}\, \frac{1}{2}}_{\pi N}-
	 \sqrt{\frac{2}{3}} \,\ket{\frac{1}{2}\, \frac{1}{2}}_{\pi N} ,	   
\end{array}
\right.
\end{eqnarray*}

\begin{eqnarray*}	 
\left\{
\begin{array}{ll}
\ket{\pi^- p} = 
         \frac{1}{\sqrt 3}  \,\ket{\frac{3}{2}\,-\frac{1}{2}}_{\pi N} -
         \sqrt{\frac{2}{3}} \,\ket{\frac{1}{2}\,-\frac{1}{2}}_{\pi N} , \\
%
%
\ket{\pi^0 n} =
         \sqrt{\frac{2}{3}} \,\ket{\frac{3}{2}\,-\frac{1}{2}}_{\pi N} +
         \frac{1}{\sqrt 3}  \,\ket{\frac{1}{2}\,-\frac{1}{2}}_{\pi N} ,
\end{array}	  
\right.
\end{eqnarray*}

\noindent
and the following single state:

\begin{eqnarray*}
\ket{\pi^- n} = \ket{\frac{3}{2}\,-\frac{3}{2}}_{\pi N} .
\end{eqnarray*}

Retaining only the contributions with total isospin 3/2, we deduce
from the above expressions the following relations between the strength 
parameters in the two bases:

\begin{eqnarray}
\lambda_{\pi^0 p - \pi^0 p } &=& \lambda_{\pi^0 n - \pi^0 n } =  
                   \frac{2}{3}\lambda_{\Delta} \quad , \quad
\lambda_{\pi^+ n - \pi^+ n } = \lambda_{\pi^- p - \pi^- p } =
		   \frac{1}{3}\lambda_{\Delta} ,  
\end{eqnarray}

\noindent
etc.

	       \subsection{$\Sigma N$-$\Lambda N$ states}

The $\Sigma N$-$\Lambda N$ states in the particle basis are expressed as 
linear combinations of the $I=\frac{1}{2}, \frac{3}{2}$ states. We have
two sets of three coupled states:

\begin{eqnarray*}
\left\{
\begin{array}{ll}
\ket{\Sigma^0 p} = 
         \sqrt{\frac{2}{3}} \,\ket{\frac{3}{2}\, \frac{1}{2}}_{\Sigma N} -
	 \frac{1}{\sqrt 3}  \,\ket{\frac{1}{2}\, \frac{1}{2}}_{\Sigma N} , \\
%
%
\ket{\Sigma^+ n} = 
       - \frac{1}{\sqrt 3}  \,\ket{\frac{3}{2}\, \frac{1}{2}}_{\Sigma N}-
	 \sqrt{\frac{2}{3}} \,\ket{\frac{1}{2}\, \frac{1}{2}}_{\Sigma N} , \\
%
%
\ket{\Lambda p} = \ket{\frac{1}{2}\, \frac{1}{2}}_{\Lambda N} , 	 
\end{array}
\right.
\end{eqnarray*}

\begin{eqnarray*}	 
\left\{
\begin{array}{ll}
\ket{\Sigma^- p} = 
         \frac{1}{\sqrt 3}  \,\ket{\frac{3}{2}\,-\frac{1}{2}}_{\Sigma N} -
         \sqrt{\frac{2}{3}} \,\ket{\frac{1}{2}\,-\frac{1}{2}}_{\Sigma N} , \\
%
%
\ket{\Sigma^0 n} = 
         \sqrt{\frac{2}{3}} \,\ket{\frac{3}{2}\,-\frac{1}{2}}_{\Sigma N} +
         \frac{1}{\sqrt 3}  \,\ket{\frac{1}{2}\,-\frac{1}{2}}_{\Sigma N} , \\
%
%
\ket{\Lambda n} = \ket{\frac{1}{2}\, -\frac{1}{2}}_{\Lambda N} . 	 
\end{array}	  
\right.
\end{eqnarray*}

\noindent
and the two following single states:

\begin{eqnarray*}
\ket{\Sigma^- n} &=& \ket{\frac{3}{2}\,-\frac{3}{2}}_{\Sigma N} ,
\end{eqnarray*}

\begin{eqnarray*}
\ket{\Sigma^+ p} &=& \ket{\frac{3}{2}\, \frac{3}{2}}_{\Sigma N} .
\end{eqnarray*}

From these relations, we can express the strength parameters in the particle
basis in terms of those in the isospin basis. For example, we have:

\begin{eqnarray}
\lambda_{\Sigma^0 p - \Sigma^+ n } &=& -\lambda_{\Sigma^- p - \Sigma^0 n} =  
     \frac{\sqrt 2}{3}[ \lambda_{\Sigma N-\Sigma N}^{1/2} -
                        \lambda_{\Sigma N-\Sigma N}^{3/2} ] 
      \quad , \quad ,
\lambda_{\Sigma^+ n - \Lambda p } = \lambda_{\Sigma^- p - \Lambda n} =
		   - \sqrt \frac{2}{3}\lambda_{\Sigma N-\Lambda N}^{1/2} , 
\end{eqnarray}
  
\noindent
and so on.

\section{}
\protect\label{app-b}

\centerline{\bf Separable model for coupled channels}

The Bethe-Salpeter equation \cite{Sal51} is the relativistic 
generalization of the Lippmann-Schwinger equation describing the scattering 
of two particles. In the case of $n$ coupled two-body channels, we have 
a system of $n$ coupled equations, which reads, in operator form~:

\begin{equation}
\label{eq:eq2c-tij}
T_{ij}(\sigma) = V_{ij} + \sum_{k} V_{ik} G_0^k(\sigma) T_{kj}(\sigma) ,
\end{equation}

\noindent
with $\sigma$ the square of the total centre of mass energy. 
The indices $\{i,j,k\}$ run over the $n$ 
two-body channels. $G_0$ is the two-body propagator, $V_{ij}$ the 
transition potential between channel $i$ and $j$, and $T_{ij}$ the 
$t$-matrix for that transition.

After projecting in the 4-momentum space representation, we obtain
a system of 4-dimensional integral equation. 
Using the Blankenbecler-Sugar (BbS) method (Refs. \cite{Bla66,Aar77,Gir78}), 
this equation can be reduced to 
the following set of coupled three-dimensional equations: 

\begin{equation}
\label{eq:t_ij}
T_{ij}(\bm p_i , \bm p_j ; \sigma) =  
V_{ij}(\bm p_i , \bm p_j ; \sigma) +
  \sum_{k}\int d^3 p_k V_{ik}(\bm p_i , \bm p_k; \sigma)
            G_0^k(p_k;\sigma) T_{kj}(\bm p_k , \bm p_j ; \sigma) ,
\end{equation}	

\noindent
where $\bm p_i$ is the momentum in channel $i$. The relativistic two-body
propagator has the following relativistic expression:

\begin{equation}
\label{eq:G0_2c}
G_0^k(p_k;\sigma) = \frac{1}{(2\pi)^3}
           \frac{\epsilon_1+\epsilon_2}
                {2\epsilon_1\epsilon_2 [ (\epsilon_1+\epsilon_2)^2 -\sigma ]} ,
\end{equation}

\noindent
where the two particles in channel $k$ are labeled as 1 and 2, and
$\epsilon_i=\sqrt {\bm p_i^{\>2}+m_i^2}$ is the energy of particle~$i$ 
with mass $m_i$. 

Finally, in the case of $s$-wave interactions, Eq. (\ref{eq:t_ij}) reduces
to the following set of coupled one-dimensional integral equations:

\begin{equation}
\label{eq:t2c_swave}
T_{i j}(p_i , p_j ; \sigma) =  V_{i j}(p_i , p_j; \sigma)
  + \sum_{k}\int_0^{\infty} p_k^2dp_k V_{i k}(p_i , p_k; \sigma)
            G_0^k(p_k;\sigma) T_{k j}(p_k , p_j ; \sigma) . 
\end{equation}

Now, we assume that $V_{ij}$'s are separable, i.e. we write:

\begin{equation}
\label{eq:vij-sep}
V_{i j}(p_i , p_j; \sigma) = g_{i}(p_i)\, \lambda_{i j}(\sigma)\, g_{j}(p_j) .
\end{equation}

\noindent
The $g$'s are the form factors, and $\lambda_{i j}(\sigma)$ is the strength 
for the transition $i \leftrightarrow j$.  Here we assume that in general  
$\lambda$'s  are functions of $\sigma$. Using this expression in
Eq. (\ref{eq:t2c_swave}), we obtain $T$ as separable:

\begin{equation}
\label{eq:tij-sep}
T_{i j}(p_i , p_j ;\sigma) = g_{i}(p_i) R_{i j}(\sigma) g_{j}(p_j) , 
\end{equation}

\noindent
where $R_{i j}$ is an element of the following ($n \times n$) matrix:

\begin{equation}
\label{eq:rs}
R(\sigma) = [\lambda^{-1}(\sigma) - \widetilde G(\sigma) ]^{-1} .
\end{equation}

\noindent
Here $\lambda(\sigma)$ is an $n\times n$ matrix of the strengths, 
and $\widetilde G(s)$ is a diagonal 
matrix with the following elements:

\begin{equation}
\label{eq:gijs}
\widetilde G_{i j}(\sigma) = \delta_{i j} 
             \int_0^{\infty} p^2dp\, g_i^2(p)\, G_0^{\,i}(p;\sigma) , 
\end{equation}

\noindent
where $G_0^{\,i}$ is the relativistic two-body propagator for channel $i$,
calculated from Eq. (\ref{eq:G0_2c}). 

Note that in the case where the strength matrix has no inverse, 
Eq. (\ref{eq:rs}) must be re-written as:

\begin{equation}
\label{eq:rs-bis}
   R(\sigma) = [\bm 1-
        \lambda(\sigma) \widetilde G(\sigma)]^{-1} \lambda(\sigma) , 
\end{equation}  

\noindent
where $\bm 1$ is the ($n \times n$) unit matrix.
This situation occurs for example when the $\Kbar N$ interactions are 
considered in the particle basis, as the strength parameters obtained as 
linear combinations of the isospin basis values constitute a matrix
with its determinant equal to zero.

\section{}
\protect\label{app:antisym}

\centerline{\bf Antisymmetrization}

In the isospin basis, the two nucleons are considered as identical particles.
Therefore, one must construct Born terms and three-body amplitudes
properly antisymmetrized with respect to the two nucleons.
We present hereafter the principle of the method, and we refer the reader
to Refs. \cite{Bah_thesis,Gir78} for more details. 

Let us label the two nucleons as $N_1$ and $N_2$. 
The $d$ channel corresponds to $\Kbar(\widetilde{N_1N_2})$, where the tilde
means that the deuteron wave function is properly antisymmetrized.
We must also add the nucleons labels to the labels defined 
in Table \ref{tab:labels}:  
$\Delta^i=\Sigma(\pi N_i)$, $\beta^i= \pi(Y N_i)$, $\alpha^i= N_i(\pi Y)$,
with $i=1,2$, and $y^i= N_j(\Kbar N_i)$, with $i,j=1,2$ $(i\ne j)$.
(note that the lower indices have been removed, since they do not participate
in this discussion).  

Now, we have two types of Born term, depending on whether one or two nucleons
are involved. In the first category, we have $Z_{\Delta \alpha}$,
$Z_{\Delta \beta}$, and  $Z_{\alpha \beta}$. For these terms, we only
need to introduce the nucleon index. For example, we define:
$Z_{\Delta^1 \alpha^1}=\bra{\Sigma(\pi N_1)} G_0 \ket{N_1(\pi \Sigma)}$,
$Z_{\Delta^2 \alpha^2}=\bra{\Sigma(\pi N_2)} G_0 \ket{N_2(\pi \Sigma)}$.,
where $G_0$ is the three-body propagator.
Since nucleons 1 and 2 are identical, it is clear that these two Born terms
are identical, and we will set:
$\widetilde{Z}_{\Delta \alpha}=Z_{\Delta^1 \alpha^1}=Z_{\Delta^2 \alpha^2}$.
In the same way, we define: 
$\widetilde{Z}_{\Delta \beta}=Z_{\Delta^1 \beta^1}=Z_{\Delta^2 \beta^2}$,
and:
$\widetilde{Z}_{\alpha \beta}=Z_{\alpha^1 \beta^1}=Z_{\alpha^2 \beta^2}$.

The Born terms involving two nucleons, namely $Z_{dy}$ and $Z_{yy}$, must be
antisymmetrized. As the $d$ state is already antisymmetric, we only need 
to antisymmetrize the $y$ state. So, using the notations defined above, 
the antisymmetric $Z_{dy}$ Born term will be defined as:
       
\begin{equation}
\widetilde{Z}_{dy}= \bra{\Kbar(\widetilde{N_1N_2})} G_0 
   \frac{1}{\sqrt 2}\left [ \ket{N_2(\Kbar N_1)} - \ket{N_1(\Kbar N_2)}\right] =
   \frac{1}{\sqrt 2} [ Z_{dy^1} - Z_{dy^2} ] .
\end{equation} 

\noindent
Exchanging $N_1$ and $N_2$ in  $Z_{dy^1}$, it is obvious that: 
$Z_{dy^1} = - Z_{dy^2}$, thus:

\begin{equation}
\widetilde{Z}_{dy}= {\sqrt 2}  Z_{dy^1} . 
\end{equation} 

We proceed along the same lines to define the antisymmetric $Z_{yy}$ 
Born term:

\begin{equation}
\widetilde{Z}_{yy}= 
\frac{1}{\sqrt 2} \left [ \bra{N_2(\Kbar N_1)} - \bra{N_1(\Kbar N_2)} \right]
          G_0 
\frac{1}{\sqrt 2} \left [ \ket{N_2(\Kbar N_1)} - \ket{N_1(\Kbar N_2)} \right] . 
\end{equation}   

\noindent
As $\Kbar$ can be exchanged only between the $(\Kbar N_1)$ and $(\Kbar N_2)$
pairs, we have:

\begin{equation}
\widetilde{Z}_{yy} = -\frac{1}{2} 
\left [ \bra{N_2(\Kbar N_1)} G_0 \ket{N_1(\Kbar N_2)}  +
       \bra{N_1(\Kbar N_2)} G_0 \ket{N_2(\Kbar N_1)} \right] =
   \frac{1}{2} [ Z_{y^1y^2} + Z_{y^2y^1} ] .
\end{equation}

\noindent
Due to the identity of the two nucleons, we have: $Z_{y^1y^2} = Z_{y^2y^1}$,
and thus:

\begin{equation}
\label{eq:zyy}
\widetilde{Z}_{yy}= - Z_{y^1y^2} . 
\end{equation} 

For the practical calculation, only the coefficients appearing in the
expressions of the antisymmetric Born terms are important, and we can 
ignore the nucleons labels.

Concerning the two-body propagators, the nucleons labels can be ignored,
since only one nucleon is eventually involved in the propagating pair.  
(the deuteron propagator corresponds to a properly symmetrized $^3S_1$
or $^3S_1$-$^3D_1$ state).

Finally, the three-body equations can be rewritten in the same form
as Eq. (\ref{eq:3body}), where now the Born terms are the antisymmetric
terms as defined above. 

\end{appendix}


\end{document}